\documentstyle[12pt]{article}

\newcommand{\al}{\alpha}
\newcommand{\ep}{\epsilon}

\newcommand{\la}{\lambda}
\newcommand{\La}{\Lambda}

\newcommand{\lb}{\lbrack}
\newcommand{\rb}{\rbrack}

\newcommand{\msc}[1]{\mbox{\scriptsize #1}}
\newcommand{\dsp}{\displaystyle}

\newcommand{\br}{\mbox{{\bf R}}}
\newcommand{\bz}{\mbox{{\bf Z}}}

\newcommand{\bsz}{\msc{{\bf Z}}}

\newcommand{\bu}{\mbox{{\bf u}}}

\newcommand{\cG}{{\cal G}}
\newcommand{\cJ}{{\cal J}}
\newcommand{\cT}{{\cal T}}
\newcommand{\cO}{{\cal O}}
\newcommand{\cN}{{\cal N}}
\newcommand{\cD}{{\cal D}}

\newcommand{\ket}[1]{{|#1\rangle}}
\newcommand{\bra}[1]{{\langle#1|}}

\newcommand{\tr}{\mbox{Tr}}

\newcommand{\tG}{\tilde{G}}
\newcommand{\tL}{\tilde{L}}
\newcommand{\tJ}{\tilde{J}}
\newcommand{\tT}{\tilde{T}}
\newcommand{\tq}{\tilde{q}}
\newcommand{\tj}{\tilde{j}}
\newcommand{\tA}{\tilde{A}}
\newcommand{\tpsi}{\tilde{\psi}}
\newcommand{\tphi}{\tilde{\phi}}

\newcommand {\eqn}[1]{(\ref{#1})}

\newcommand{\nn}{\nonumber\\}

\def\n{\nonumber}
\def\R{{\bf R}}
\def\Z{{\bf Z}}
\def\C{{\rm Ch}}

\def\o{\over}

\makeatletter
\@addtoreset{equation}{section}
\def\theequation{\thesection.\arabic{equation}}
\makeatother

\begin{document}

\vskip 7mm
\begin{titlepage}
 
 \renewcommand{\thefootnote}{\fnsymbol{footnote}}
 \font\csc=cmcsc10 scaled\magstep1
 {\baselineskip=14pt
 \rightline{
 \vbox{\hbox{hep-th/0101211}
       \hbox{UT-922}
       }}}

 \vfill
 \baselineskip=20pt
 \begin{center}
 \centerline{\Huge  Formation of Spherical D2-brane} 
 \vskip 5mm 
 \centerline{\Huge from Multiple D0-branes}

 \vskip 2.0 truecm

\noindent{\it \large Yasuaki Hikida, Masatoshi Nozaki  and Yuji Sugawara} \\
{\sf hikida@hep-th.phys.s.u-tokyo.ac.jp~,~
nozaki@hep-th.phys.s.u-tokyo.ac.jp~,~\\ sugawara@hep-th.phys.s.u-tokyo.ac.jp}
\bigskip

 \vskip .6 truecm
 {\baselineskip=15pt
 {\it Department of Physics,  Faculty of Science\\
  University of Tokyo\\
  Bunkyo-ku, Hongo 7-3-1, Tokyo 113-0033, Japan}
 }
 \vskip .4 truecm

 \end{center}

 \vfill
 \vskip 0.5 truecm

\begin{abstract}
\baselineskip 6.7mm
We study D-branes in $SU(2)$ WZW model by means of the boundary state
techniques. We realize the ``fuzzy sphere'' configuration of multiple 
D0-branes as the boundary state with the insertion of suitable Wilson line. 
By making use of the path-integral representation   we show that 
this boundary state preserves the appropriate boundary conditions
and leads to the Cardy state describing a spherical D2-brane 
under the semi-classical approximation. This result directly 
implies that the spherical D2-brane in $SU(2)$ WZW model 
can be well described as the bound state 
of D0-branes.

After presenting  the supersymmetric extension,  we also 
investigate the BPS and the non-BPS configurations of D-branes 
in the NS5 background. 
We demonstrate that the non-BPS configurations are actually unstable,
since they always possess the open string tachyons. We further notice
that the stable BPS bound state constructed  by the tachyon condensation
is naturally interpreted as the brane configuration of fuzzy sphere.
\end{abstract}

\setcounter{footnote}{0}
\renewcommand{\thefootnote}{\arabic{footnote}}
\end{titlepage}

\newpage

\section{Introduction}

Toward thorough  understanding of D-brane dynamics,
the studies on the flat D-branes with constant $B$-field have  
recently received a great deal of interests \cite{cds,dh,sw}. 
One of the important precepts of them is the existence 
of two equivalent descriptions of such backgrounds; 
one is the ``commutative description'', of which 
low energy effective action is well-known DBI action, and the other  
is the ``non-commutative description'' based on the geometry 
induced by the open string vertices.
A transparent viewpoint for the understanding of 
appearance of non-commutativity is 
the interpretation of D-branes with constant $B$-field 
as the bound states of infinitely many lower dimensional D-objects 
(D0-branes or D-instantons, typically). 
This has been a well-known aspect in the context of Matrix theory
\cite{bfss,ikkt}.
Roughly speaking, the classical solution of Matrix theory 
corresponding to the ``non-commutative configuration''
\begin{equation}
 \lb X^i,~ X^j \rb = i \theta^{ij}~,
\label{nc flat}
\end{equation}  
leads to such a bound state. This aspect was also discussed from the
viewpoints of perturbative string theory in the papers
\cite{ishibashi,kuroki}.  In those works the matrix coordinates $X^i$ 
are  naturally identified with the CP factors and 
are incorporated  appropriately as the (T-dualized) Wilson line such as
$\dsp \tr \left(P \exp\left(i\int d\sigma P_i(\sigma) X^i\right)\right)$.

A simple extension of these non-commutative descriptions of D-branes
to the {\em curved\/} background is realized by considering 
the {\em linear\/} $B$-field (namely, the constant field strength $H=dB
= \mbox{const} \neq 0$) rather than the constant $B$-field;
\begin{equation}
 \lb X^i,~ X^j \rb = i\ep^{ijk}X^k~, ~~~(i,j,k=1,2,3)~,
\label{fuzzy sphere}
\end{equation} 
which is the configuration called the ``fuzzy sphere'' \cite{hop}.

It is known that the fuzzy sphere configuration of D0-branes  
\eqn{fuzzy sphere} 
is a classical solution under  the suitable RR flux 
and describes the spherical brane \cite{myers,tr}, often called 
``giant graviton'' \cite{giant}. 
On the other hand, there have been many studies on the various aspects
of D-branes in $SU(2)$ WZW model,
which has a constant NS field strength $H\neq 0$ 
\cite{dwzw,alek1,alek2,bds}. 
Among other things, it was shown in \cite{bds}
that the spherical D2-branes in $SU(2)$ WZW model  
are stabilized by the flux of D0-brane charges\footnote
      {Related works about the quantization of flux on the
       spherical D2-brane are also given in \cite{taylor}.}.
It was also claimed  \cite{alek1,alek2} that the spherical D2-branes are 
no other than the bound states of D0-branes.
This claim is based on the low energy effective field theory 
in the ``non-commutative description'' of D-branes in WZW model, 
which is given as an analog of non-commutative Yang-Mills theory 
with Chern-Simons term \cite{alek2}.
Related works treating such a non-commutative gauge  theory have been given
in the recent literatures \cite{iso,hk}.

The main purpose of this paper is to clarify such a proposed 
aspect with respect to the ``formation of spherical branes'' 
from the picture of boundary states.
Starting from  the boundary state description of D0-brane in
the background of $SU(2)$ WZW model, we realize the several configurations 
of multiple D0-branes by inserting the  Wilson lines
following the works \cite{ishibashi,kuroki}. 
Especially, we focus on the fuzzy sphere configuration
of D0-branes in the similar manner  as the configuration of
non-commutative $\br^{2r}$ (or  non-commutative torus) \eqn{nc flat} 
in the case of flat string background.
We prove that such a boundary state with the Wilson line 
of fuzzy sphere satisfies the appropriate boundary condition
using the path-integral representation.
We further demonstrate how it reduces to the boundary state 
describing the stable spherical D2-brane under 
the large $k$ limit.

One of the most important applications of these results to superstring
theory is surely the study of D-branes in the NS5 background, 
since the near horizon geometry  of this background is known 
to be described by the CHS $\sigma$-model \cite{chs}. 
We consider the type II string theory in this background 
and investigate various configurations of multiple D0-branes 
with the world-sheet superconformal symmetry preserved.  
We show that, among these on-shell configurations of 
D0-branes, only the configuration of fuzzy sphere is BPS, 
and the other non-BPS configurations always 
include  tachyonic excitations in the open string spectrum. 
This means that the system of multiple D0-branes is unstable 
and  decays into a single BPS D2-brane  described by the CP factor  
of fuzzy sphere after the open string tachyons condensate. 
This aspect is reminiscent of the tachyon condensation 
in the various systems of non-BPS brane configurations  
under  the flat string background \cite{tachyon}, 
and  seems to fit with the observation by \cite{alek1,alek2} 
at the level of low energy effective field theory. The other studies 
of D-branes in the NS5 background from different viewpoints
are given in \cite{giveon,lerche,pelc,eguchi}.

~

This paper is organized as follows. In section 2  we study D-branes in 
bosonic $SU(2)$ WZW model. We examine  various configurations of multiple 
D0-branes by inserting the Wilson line operators and explain how the fuzzy
sphere configuration of D0-branes leads to a spherical D2-brane wrapped
on a conjugacy class. 
In section 3, after providing the extension to the superstring case,
we especially  study  whether or not the brane configuration preserves
the space-time SUSY and investigate the stability of the system, namely, the 
absence/existence of the tachyons in the open string spectrum.  We also 
discuss the formation of the spherical brane from 
the viewpoints of tachyon condensation.   
Section 4 is devoted to a summary and some discussions.


~


\section{Spherical D2-brane from D0-branes in Bosonic $SU(2)$ WZW model}

\subsection{D-branes in Bosonic $SU(2)$ WZW model}

We shall begin with a brief review about D-branes in
$SU(2)$ WZW model, which are  established  in  many works \cite{dwzw}.
In WZW model there are two currents; one is the left mover and the
other is the right mover,
\begin{equation}
  J^a(\tau,\sigma) = \sum _n J^a_n e^{-in ( i \tau + \sigma )}~,~~
 \tJ^a(\tau,\sigma) = \sum _n  \tJ^a_n e^{in ( -i  \tau + \sigma ) }~,
\end{equation}
where $a$ takes the values $1,2,3$ and $n$ is an integer.
D-branes should be  described by the boundary of the string world-sheet.
We have to impose an appropriate boundary  condition\footnote
    {Throughout this paper we shall write down the boundary conditions
       in the closed string channel.}. 
In maximally symmetric case, this boundary condition reads as 
\begin{equation}
 ( J^a_n + \La^a_{~b}\tJ^b_{-n}) \ket{B} = 0~,
\label{glue0}
\end{equation}
where $\La$ is an automorphism of $SU(2)$. If this is an inner 
automorphism, we can readily reduce this condition  to the simplest one  
\begin{equation}
 ( J^a_n + \tJ^a_{-n}) \ket{B} = 0 ~,
\label{glue1}
\end{equation}
by considering the rotation of the type 
$\ket{B}~\longrightarrow ~e^{iu^a\tJ^a_0}\ket{B}$.
The case when $\La$ is an outer automorphism 
is quite interesting and has been attracting much attentions
in the recent works \cite{fs,pb}. 
However, it is beyond the scope of this paper
and we shall start with the simplest ``gluing condition'' 
\eqn{glue1}.
We must also impose the conformal invariance at the boundary;
\begin{equation}
  ( L_n - \tL_{-n}) \ket{B} = 0 ~.
\label{glue2}
\end{equation}  
However, since the energy-momentum tensor is given as 
the quadratic form of currents,
this condition  is automatically satisfied 
because of the gluing condition for the currents
\eqn{glue0} or \eqn{glue1}. 

For convenience we define the following currents at the boundary with setting
$\tau = 0$,  
\begin{equation}
 J^a_{\pm} (\sigma) = \sum_n (J^a_n \pm \tJ^a_{-n}) e^{-in\sigma}~.
\end{equation}
These currents satisfy the commutation relations
\begin{eqnarray}
 {[} J^a _{\pm} (\sigma_1) ,  J^b _{\pm} (\sigma_2) {]} &=&
   2 \pi i \epsilon ^{ab}_{~~c} J^c _+ (\sigma_1) \delta ( \sigma_1 - \sigma_2)
   \label{Jpm comm 1}  ~, \\
 {[} J^a _{\pm} (\sigma_1) ,  J^b _{\mp} (\sigma_2) {]} &=&
   2 \pi i \epsilon ^{ab}_{~~c} J^c _- (\sigma_1) \delta ( \sigma_1 - \sigma_2)
  + 2 \pi i k \delta^{ab} \delta ' (\sigma_1 - \sigma _2) ~.
  \label{Jpm comm 2}
\end{eqnarray}
Here we use $'$ as the derivative with respect to $\sigma_1$.
By using these currents and the ``boundary energy-momentum tensor''
\begin{equation}
 T_- (\sigma) = \sum_n (L_n - \tL_{-n}) e^{-in \sigma}~,
\label{bemt}
\end{equation}
we can rewrite the gluing conditions \eqn{glue1}, \eqn{glue2} as
\begin{eqnarray}
 J^a _+ (\sigma) \ket{B} &=& 0~, \label{glue3}\\ 
 T_-  (\sigma) \ket{B} &=& 0~.
\end{eqnarray}
Since the currents $J^a_+(\sigma)$ generate the adjoint action of 
$SU(2)$ at the boundary, the gluing condition \eqn{glue3} 
roughly means that the directions along the conjugacy class should obey
the Neumann boundary condition, which is an origin
of the geometrical interpretations of the boundary states of interest.  

A complete system of solutions of the gluing conditions \eqn{glue1},
\eqn{glue2} is given in \cite{ishi} and called ``Ishibashi states''.
The Ishibashi states $\ket{\ell}_I$ 
($ \ell = 0,\ldots, k$) are characterized by the relation
\begin{equation}
 {}_{I}\bra{\ell} \tq^{ H^{(c)}}
  \ket{{\ell}'}_I 
 = \delta_{\ell {\ell}'} \chi^{(k)}_{\ell} (\tq)~, 
\end{equation} 
where $\dsp H^{(c)}\equiv \frac{1}{2}(L_0 + \tL_0 - \frac{c}{24})$
denotes the Hamiltonian in the closed string channel.
Here $c$ is the central charge of the system 
and  $\tq = \exp(- 2 \pi i / \tau )$ is the modulus in 
the closed string channel.  Later we will use 
$q = \exp(2 \pi i \tau )$ as the open string one.  
Furthermore $\chi^{(k)}_{\ell} (\tq)$ denotes  the $SU(2)$ character
and its modular transformation is given by
\begin{equation}
 \chi^{(k)}_{\ell} (\tq) = \sum_{{\ell}'} S_{\ell {\ell}'} 
  \chi^{(k)}_{{\ell}'} (q) ~,
\end{equation}  
where we set  
\begin{equation}
 S_{\ell {\ell}'} = \sqrt{\frac{2}{k+2}}
 \sin \left(\pi \frac{(\ell+1)({\ell}' +1)}{k+2}\right)~.
\end{equation}

Because of the linearity of the gluing conditions,
arbitrary linear combinations of Ishibashi states also satisfy them.
Among them it is convenient in some physical reasons to take
the ``Cardy states'' \cite{cardy} defined by  
\begin{equation}
 \ket{L}_C = \sum_{\ell} \frac{S_{L \ell}}{\sqrt{S_{0 \ell}}} \ket{\ell}_I ~,
\label{cardy}
\end{equation}  
where the label $L$ takes  also the values  $L = 0,1, \cdots , k$. 
It is well-known \cite{dwzw} that the each Cardy state corresponds to
the D-brane wrapped on  the  conjugacy class located at 
the quantized azimuthal angle $\dsp \theta=\frac{\pi L}{k}$, which has 
the spherical topology and is centered at the origin $e$ (the identity
element of $SU(2)$). The special example $L=0$ (and also $L=k$) 
corresponds to the point-like conjugacy class,
and thus it is naturally identified with the boundary state of D0-brane;
\begin{equation}
 \ket{D0}_C = \sum_{\ell} \frac{S_{0 \ell}}{\sqrt{S_{0 \ell}}} \ket{\ell}_I 
   =  \sum_{\ell} \sqrt{S_{0 \ell}} \ket{\ell}_I~.
\label{d0}
\end{equation}
The other Cardy states $L\neq 0, k$ correspond to spherical D2-branes 
stabilized by the $U(1)$ fluxes  $\dsp \int F= L+1$ \cite{bds};
\begin{equation}
 \ket{D2;L}_C =  
     \sum_{\ell} \frac{S_{L \ell}}{\sqrt{S_{0 \ell}}} \ket{\ell}_I ~.
 \label{d2}
\end{equation}



\subsection{Multiple D0-branes on $SU(2)$ WZW Model and Wilson Line}

Let us consider a configuration of $(L+1)$ D0-branes.
As is well-known, multiple D-branes are described by 
the degrees of freedom of the CP indices. 
Following  the discussion in the flat background \cite{ishibashi,kuroki},
we shall start with the boundary state of D0-branes with the 
Wilson line inserted in order to incorporate the CP degrees of freedom;
\begin{equation}
 \ket{D0;\{ M^a \}}_C = 
  \tr \left( P \exp \left( - \frac{i}{k} 
   \int^{2 \pi} _ 0 d \sigma J^a _- ( \sigma ) M^a(\sigma) \right)\right)
   \ket{D0}_C~,
\label{original}
\end{equation}  
where $P$ indicates the path-order and the coefficients of matrices are
chosen for convenience. 
We express the CP factors as $(L+1) \times (L+1)$ 
hermitian matrices $M^a$ ($a=1,2,3$).  
For example, in the simplest case $M^a = 0_{(L+1) \times (L+1)}$,
we have 
\begin{equation}
\ket{D0;\{ M^a \}}_C = (L+1) \ket{D0}_C~,
\end{equation}
which merely stands for a stack of $(L+1)$ D0-branes at the origin $e$.

Slightly more  non-trivial example is the case that $M^a$ are 
constant matrices which mutually commute with each other. 
In that case  we can eliminate the path-order symbol 
and the boundary state reduces to 
\begin{equation}
 \ket{D0;\{ M^a \}}_C = \tr \left(
\exp\left( -\frac{2\pi i}{k}J^a_{-,0}M^a\right) \right)
                         \ket{D0}_C ~.
\label{commutative}
\end{equation} 
This is nothing but a linear combination of  $(L+1)$  D0
boundary states, each of which is characterized by the gluing condition
of the general type \eqn{glue0}. The simultaneous eigenvalues of $M^a$
roughly express the positions of D0-branes on $S^3$.
Since the Wilson line factor only includes  the zero-modes of $SU(2)$ 
currents, it is obvious that this boundary state preserves 
the conformal invariance at the boundary.

It is more interesting to consider the example 
with the constant matrices $M^a$ satisfying
the commutation relation
\begin{equation}
 {[} M^a , M^b {]} = i \epsilon^{ab}_{~~c} M^c~,
\label{fuzzy CP}
\end{equation}
which defines the fuzzy sphere configuration of D0-branes \cite{hop}.  
In view of the conjecture given in \cite{alek1,alek2}
one will expect that this configuration leads to the spherical 
D2-brane as the bound state of $(L+1)$ D0-branes in the analogous 
way as the discussion \cite{ishibashi} in the flat background.
From now on,  we shall focus on this configuration and will show that 
this expectation is indeed the case at least under the large $k$ limit.

Since our CP matrices \eqn{fuzzy CP} are non-commutative, the
path-ordering  in the Wilson line essentially contributes 
and makes it  difficult to confirm whether or not the 
boundary state \eqn{original} satisfies the appropriate gluing condition.
At this point it is more convenient to rewrite the path-ordered trace
by means of the path-integral representation 
\cite{clny,ishibashi,kuroki,okuyama}.
The formula of path integral on a group manifold is given in
\cite{stone} (see also \cite{wiegmann}) and we apply it to our case.
The Wilson line  hence can be represented (up to normalization) as
 \begin{equation}
  \tr \left( P \exp \left( - \frac{i}{k} 
   \int^{2 \pi} _ 0 d \sigma J^a _- ( \sigma ) M^a \right)\right)
    = \int \cD g \exp \left[  -A  \int ^{2 \pi} _{0} d \sigma 
  \bra{g(\sigma)} D^{J_-}_{\sigma} \ket{g(\sigma)}\right] ~,
\label{path integral}
\end{equation} 
where we set 
\begin{equation}
  D^{J_-}_{\sigma} = 
 \frac{d}{d \sigma} + \frac{i}{k} J^a _- (\sigma) M^a ~.
\end{equation}
The integral variable $g(\sigma)$ 
is  the map of $S^1$ to $SU(2)$ and we  define the ``coherent state'' as 
\begin{equation}
 \ket{g(\sigma)} = R_L ( g(\sigma))\ket{0} ~, ~~ 
 \bra{g(\sigma)} = \bra{0}R_L ( g(\sigma)^{-1}) ~,
\end{equation}
where $R_L$ denotes the spin $L/2$ representation of $SU(2)$
and $\ket{0}$, $\bra{0}$ denote the highest weight vector and its dual.
We can  also suppose  that $M^a=R_L(T^a)$ without loss of generality, 
where $T^a$ denote the generators of $SU(2)$ algebra. 
The factor $A$ is nothing but a normalization constant and we can prove
that the path-integral does not depend on this constant (up to the
overall normalization), which will be explained just below.

Precisely speaking, we have to perform the gauge fixing about 
the gauge symmetry $g(\sigma)\, \longrightarrow \, g(\sigma)h(\sigma)$
$(\forall h(\sigma)\in U(1))$ in order to define the path-integral
properly. The reduced phase space, which has the topology 
$S^2 \cong SU(2)/U(1)$, is identified with the co-adjoint orbit of 
$SU(2)$ and thus canonically equipped with the Kirillov-Kostant 
symplectic structure. This can be also identified with the conjugacy 
class located at $\dsp \theta = \frac{\pi L}{k} $. This aspect is  
quite expected for our later discussion 
and is similar to that for the flat background \cite{ishibashi},
in which the phase space where the path-integral is carried out 
is naturally identified with the world-volume of the brane created
as a bound state.

The canonical quantization on the reduced phase space
(so-called the ``geometric quantization'' \cite{woodhouse}) 
provides  the finite dimensional quantum Hilbert space
naturally identified with the representation space of $SU(2)$
with spin $L/2$. The quantum mechanical operators, which are
necessarily finite dimensional matrices, are important objects 
in the non-commutative description of spherical brane. 
The finite dimensionality of them  is the origin of the fuzziness of 
spherical brane as was pointed out in many literatures. 
To be more specific, the function $m^a(g)\equiv \bra{g}M^a\ket{g}$
defines an observable on the reduced phase space, since this is 
$U(1)$ gauge invariant, $m^a(gh)=m^a(g)$, ($\forall h \in U(1)$).
According to the argument of \cite{stone}, we can show that
\begin{eqnarray}
\{m^a,~m^b\}_{\msc{PB}}&=&\frac{1}{A}\bra{g}\lb M^a,~M^b\rb \ket{g} \nn
&=&\frac{1}{A}i\ep^{ab}_{~~c}\bra{g} M^c\ket{g}
=\frac{1}{A}i\ep^{ab}_{~~c}m^c~.  
\label{PB}
\end{eqnarray}
This implies that the quantum mechanical operator $\hat{m}^a$
corresponding to $m^a(g)$ should be identified with the matrix 
$\dsp \frac{1}{A}M^a$. We here  point out that this factor $1/A$
cancels the overall factor $A$ in the path-integral formula   
\eqn{path integral}, which proves that the path-integral 
does not actually depend on the choice of $A$. 
From now on, we shall set $A=1$, which is the convention
taken in \cite{stone}.

We should also remark that it is not manifest whether this path-integral 
is well-defined or not, since the non-commutative 
operators $J^a_-(\sigma)$ appear  in the integrand. 
Nevertheless we can safely use this representation because  
the terms  $J^a _-(\sigma) \bra{g(\sigma)}M^a\ket{g(\sigma)}$ and 
$J^a _-(\sigma') \bra{g(\sigma')}M^a\ket{g(\sigma')}$
commute with each other for an  arbitrary choice of $\sigma$, $\sigma'$
and thus we can deal with it just like a c-number. 
In this evaluation of commutator it is crucial that currents $J^a_-$
have no Schwinger term. (See \eqn{Jpm comm 1}.) 


Now we show that the boundary state \eqn{original} with 
the CP factor of fuzzy sphere 
satisfies the gluing conditions \eqn{glue1}, \eqn{glue2}.
To this end it is enough to prove that the Wilson line \eqn{path
integral} commutes with all the currents $J^a_+(\sigma)$  and the boundary 
stress tensor $T_-(\sigma)$. 

Because of the commutation relations
\begin{eqnarray}
  {[} J^a _{+} (\sigma_1) ,  J^b _{-} (\sigma_2) {]} &=&
   2 \pi i \epsilon ^{ab}_{~~c} J^c _- (\sigma_1) \delta ( \sigma_1 - \sigma_2)
  + 2 \pi i k \delta^{ab} \delta ' (\sigma_1 - \sigma _2) ~, \nn
   {[} M^a , M^b {]} &=&
  i \epsilon ^{ab}_{~~c} M^c
  ~,
\end{eqnarray}
we can easily derive the equality
\begin{equation}
 \hat{U} J_- (\sigma)  \hat{U}^{-1} = 
 U(\sigma)  J_- (\sigma)  U(\sigma)^{-1} 
 - i k U(\sigma) \frac{d}{d \sigma}  U(\sigma)^{-1}~,
\end{equation}
or equivalently,
\begin{equation}
  \hat{U} D^{J_-}_{\sigma} \hat{U}^{-1} = 
 U(\sigma) D^{J_-}_{\sigma} U(\sigma)^{-1}~,
\label{DD}
\end{equation}
where we write  $J_- (\sigma) =  J_- ^a (\sigma) M^a$ and set 
\begin{equation}
 \hat{U} = \exp 
 \left( i \int^{2 \pi} _0 u^a (\sigma) J_+ ^a (\sigma) d \sigma \right) ~,~~
 U(\sigma) =  \exp ( - 2 \pi i  u^a (\sigma) M^a)~.
\label{UU}
\end{equation}
The action of  currents to the CP factors is  then written as 
  \begin{eqnarray}
 \lefteqn{ \hat{U} \, \tr \left( P \exp \left( - \frac{i}{k} 
   \int^{2 \pi} _ 0 d \sigma 
     J^a _- ( \sigma ) M^a \right)\right)  \hat{U}^{-1}} \nn
    &=& 
   \hat{U} \int \cD g \exp \left[  - \int ^{2 \pi} _{0} d \sigma 
  \bra{g(\sigma)} D^{J_-}_{\sigma} \ket{g(\sigma)}\right] \hat{U}^{-1} \nn
    &=& 
   \int \cD g \exp \left[  -  \int ^{2 \pi} _{0} d \sigma 
  \bra{g(\sigma)} U(\sigma) D^{J_-}_{\sigma} U(\sigma)^{-1} 
     \ket{g(\sigma)}\right] ~.
\label{UTU}
\end{eqnarray} 
Consider  the transformation of integral variable 
$g(\sigma) \to g'(\sigma) = U(\sigma)^{-1} g(\sigma)$.
Since the path integral measure should be invariant under 
this transformation, i.e. $\cD g'=\cD g$,
we can conclude that the Wilson line \eqn{path integral}  
is invariant under the arbitrary transformations 
generated by the  $SU(2)$ currents $J^a_+(\sigma)$,
which proves our claim.

One may be afraid that the Wilson line operator (\ref{path integral}) was
treated in  a formal way. 
Especially it would include a potential subtlety due to the UV divergence. 
However, we can define it by a suitable regularization and the above
statement is confirmed in a more rigorous way in appendix A.

As we already mentioned, the conformal invariance at the boundary
\eqn{glue2} is also satisfied.
In this way we have proved that the boundary state with 
the CP factor of fuzzy sphere
obeys the {\em same\/} gluing conditions \eqn{glue1},
\eqn{glue2}.  This fact contrasts with the cases of commutative CP
factors, in which the gluing condition \eqn{glue1} suffers the 
``twist'' into the general one \eqn{glue0}.


\subsection{Formation of Spherical D2-brane}

Now we again focus on the constant CP matrices $M^a$ 
of the fuzzy sphere \eqn{fuzzy CP}.
In the previous subsection we proved  the boundary state \eqn{original}
with  these CP matrices satisfies the same gluing condition
\eqn{glue1}. 
Since the complete system  of the solutions of 
the gluing condition \eqn{glue1} is given by 
the Ishibashi states, the boundary state of interest 
should  be represented as a linear combination of Ishibashi
states. Moreover, it is obvious that the action of Wilson line operator 
\eqn{path integral} closes in each of the  integrable module of
$\widehat{SU}(2)$. 
Therefore we can obtain the next simple relation for each of the 
Ishibashi states;
\begin{equation}
  \tr \left( P \exp \left( - \frac{i}{k} 
   \int^{2 \pi} _ 0 d \sigma J^a _- ( \sigma ) M^a \right)\right)
  \, \ket{\ell}_I = c(L,\ell,k) \ket{\ell}_I~,
\label{ishibashi wilson}
\end{equation}
where $c(L,\ell,k)$ is nothing but a c-number depending on $L$, $\ell$,
$k$. To evaluate this coefficient $c(L,\ell,k)$ 
we only have to consider the components   of primary states.
However, it is still rather difficult 
to calculate precisely the coefficient $c(L,\ell,k)$.
The best we can do now is to evaluate it under the semi-classical 
limit $k~ \rightarrow~ +\infty$. 
Naively one might suppose that we merely have 
$\dsp \tr \left( P \exp \left( - \frac{i}{k} 
   \int^{2 \pi} _ 0 d \sigma J^a _- ( \sigma ) M^a \right)\right)
\approx 1$ in this limit. But this is not correct.
We point out that in this limit 
the components of Ishibashi states with $\ell \sim k \gg 1$ 
are dominant  in the D0-brane Cardy state $\ket{D0}_C$ \eqn{d0}. 
Hence we can assume the order estimation $J^a_- \sim \ell \sim k$
and obtain the following  evaluation;
\begin{eqnarray}
\left.
\left[P \left(-\frac{i}{k}\int_0^{2\pi}d\sigma \, J^a_-(\sigma)M^a\right)^n
  \ket{D0}\right]\right|_{\msc{primary state}}
 &=& \left(-\frac{2\pi i}{k}J^a_{-,0}M^a\right)^n
\ket{D0}|_{\msc{primary state}}  \nn 
 && \hspace{3cm} + \cO \left(\frac{1}{k}\right) ~.
\label{estimation}
\end{eqnarray} 
We thus obtain in the large $k$ limit
\begin{eqnarray}
\ket{D0;\{ M^a \}}_C|_{\msc{primary}}
 &\approx& \tr  _{R_L}\left( \exp (-\frac{2\pi i}{k} J^a _{-,0} M^a)\right) 
 \ket{D0}_C |_{\msc{primary}}   \nn
 &=& \tr  _{R_L}\left( \exp (-\frac{4\pi i}{k} J^a _0 M^a)\right) 
 \ket{D0}_C |_{\msc{primary}} ~.
\label{semi}
\end{eqnarray}
Moreover, since we can now suppose $\ell \gg 1$,  
we can make use of the  semi-classical approximation for the 
``angular momentum'' such as $\dsp J_0^a \sim \sqrt{\vec{J}^2} 
n^a \sim \frac{\ell+1}{2} n^a$,
where we denote $n^a$ as a  unit vector in some direction on $S^3$.
In this approximation  the right hand side of (\ref{semi})
can be rewritten as 
\begin{eqnarray}
  \lefteqn{\tr _{R_L}\left( \exp (- \frac{4\pi i}{k} J^a _0 M^a)\right) 
 \ket{D0}_C |_{\msc{primary}}}\nn
 &\approx& 
  \sum_{\ell \gg 1} \tr _{R_L} 
 \left(\exp( - \frac{4\pi i}{k+2} \frac{\ell +1}{2} n^a M^a) \right)
\frac{S_{0 \ell}}{\sqrt{S_{0 \ell}}} \ket{\ell}_I |_{\msc{primary}}\nn
&=& \sum_{\ell \gg 1}\chi_L \left(\exp( -\frac{4\pi i}{k+2}
  \frac{\ell +1}{2} n^a M^a)\right)
\frac{S_{0 \ell}}{\sqrt{S_{0 \ell}}} \ket{\ell}_I |_{\msc{primary}}~.
 \end{eqnarray}
$\chi_L$ denotes the $SU(2)$ character of the spin $L/2$ representation.
Since this is a class function (i.e. $\chi_L(hgh^{-1})=\chi_L(g)$), 
we have
\begin{eqnarray}
 \chi_L \left(\exp( - \frac{4\pi i}{k+2}  \frac{\ell +1}{2} n^a M^a)\right) 
 &=& 
  \chi_L \left(\exp( - 2 i \pi \frac{\ell +1}{k+2} M^3)\right) \nn
 &=& \frac{\sin \left( \pi \frac{(L+1)(\ell +1)}{k+2}\right)}
     {\sin \left( \pi \frac{(\ell +1)}{k+2}\right)} ~ 
=~ \frac{S_{L\ell}}{S_{0 \ell}}~.
 \end{eqnarray}
In this way we can conclude that 
\begin{equation}
 c(L,\ell,k) \approx \frac{S_{L\ell}}{S_{0 \ell}} ,~~~(k~\rightarrow~+\infty)~.
\label{evaluation c}
\end{equation}
It leads to the remarkable result 
\begin{equation}
\ket{D0; \{M^a\}}_C \approx \ket{D2; L}_C, ~~~(k~\rightarrow~+\infty)~,
\label{D0 to D2}
\end{equation}
which  means that the fuzzy sphere configuration of $(L+1)$ 
D0-branes forms the spherical D2-brane as their bound states and this
is wrapped on the $(L+1)$-th conjugacy class of $SU(2)$.


More rigorous derivation of \eqn{evaluation c} is given as follows.
Since $\tr _{R_L}$ is taken over  the CP degrees of freedom
belonging to the spin $L/2$ representation of $SU(2)$,
we can also evaluate the \eqn{semi} in the next way;
\begin{equation}
 \tr  _{R_L}\left( \exp (- i \frac{4 \pi}{k+2} J^a _0 M^a)\right) 
 \ket{\ell}_I|_{\msc{primary}} = 
  \sum_{M =-\frac{L}{2}}^{M =\frac{L}{2}}
  \bra{L,M}  \exp (- i \frac{4 \pi}{k+2} J^a _0 M^a) 
  \ket{\ell}_I|_{\msc{primary}}
  \otimes \ket{L,M}~.
\end{equation}
Since the spin $\ell/2$ representation of 
the zero mode algebra $\{J^a_0\}$ appears in 
the Ishibashi state $\ket{\ell}_I|_{\msc{primary}}$,
we can easily diagonalize the action of the operator $J^a_0M^a$. 
The eigen-value is evaluated as 
\begin{eqnarray}
 J^a_0M^a &=&
   \frac{1}{2}\{ ( \vec{J_0} + \vec{M})^2 - \vec{J_0}^2 - \vec{M}^2 \} \nn
   &=& \frac{1}{8} \{ (\ell  - L + 2m  +1)^2 - 1  - (\ell +1)^2 +1
        - (L+1)^2 +1 \} \nn
   &=& - \frac{1}{4} (\ell +1)(L-2m) 
   + \frac{1}{8}(2m+1)^2 + \frac{1}{8}-\frac{1}{4}(L+1)(2m+1)  ~,
\label{eigen value}
\end{eqnarray}
where $m$ runs over the range $m=0,1,\cdots, L$ because of the
Clebsch-Gordan rule.
We can now assume $\ell \gg L$ because we have $\ell \gg 1$,
which implies that only the first term in \eqn{eigen value}  is dominant.
We can thus continue the evaluation  as follows;
\begin{eqnarray}
c(L,\ell,k)
  &\approx& \sum _{m=0} ^{L} \left( \exp \left(
\frac{\pi i (\ell +1)(L-2m)}{k+2}\right)\right)  \nn
 &=& \frac{\sin \left( \pi \frac{(L+1)(\ell +1)}{k+2}\right)}
     {\sin \left( \pi \frac{(\ell +1)}{k+2}\right)} 
  ~ =~ \frac{S_{L\ell}}{S_{0\ell}} ~.
\end{eqnarray}
In this way  we have got  the same result as we previously obtained.


Lastly let us discuss a simple generalization of \eqn{D0 to D2}.
Consider the boundary state 
\begin{equation}
 \ket{L_1;\{ M^a \}}_C \equiv
  \tr \left( P \exp \left( - \frac{i}{k} 
   \int^{2 \pi} _ 0 d \sigma J^a _- ( \sigma ) M^a \right)\right)
   \ket{L_1}_C~,
\label{D2 wl}
\end{equation} 
where $\{M^a\}$ are the CP matrices of fuzzy sphere 
of the size $L_2+1$. We assume $L_1,L_2 \ll k \sim +\infty $.
Then, the Ishibashi states with large $\ell$ are again dominant, and  
we can easily evaluate \eqn{D2 wl} thanks to \eqn{evaluation c}; 
\begin{eqnarray}
\ket{L_1;\{ M^a \}}_C &=& \sum_{\ell}  
  \frac{S_{L_1\ell}}{\sqrt{S_{0\ell}}}\, c(L_2,\ell, k) \ket{\ell}_I \nn
&\approx& \sum_{\ell} \frac{S_{L_1\ell}}{\sqrt{S_{0\ell}}} 
\frac{S_{L_2\ell}}{S_{0\ell}}\ket{\ell}_I \nn
&=& \sum_{\ell, L} N^L_{L_1,\,L_2} \frac{S_{L\ell}}{\sqrt{S_{0\ell}}}
\ket{\ell}_I = \sum_{L} N^L_{L_1,\,L_2} \ket{L}_C ~,
\label{D2 to D2}
\end{eqnarray}
where $N^L_{L_2,\,L_1}$ denotes the fusion matrix of
$\widehat{SU(2)}_k$ and we used  the Verlinde formula. 
This result \eqn{D2 to D2} seems to be consistent with the formula
given in \cite{alek2,fs}.

~


\section{Spherical D2-brane from D0-branes in $SU(2)$ Super WZW model}

\subsection{Preliminary}

In this section we extend the discussions in the previous
section to the supersymmetric case. It is a familiar fact that 
the near horizon physics of NS5-branes in type II string 
theory is described by the CHS $\sigma$-model \cite{chs},
$\br_{\phi}\times S^3$ in which the $S^3$ sector is 
described by $SU(2)$ super WZW model. 
Our main purpose in this section is to study the similar aspects   
of the multiple D0-branes and spherical D2-brane 
under the NS5 background in superstring theory\footnote
       {The terminologies  ``D0'', ``D2'' here are somewhat inaccurate. 
       The reader should understand that we are now focusing only on
       the sector of $SU(2)$ super WZW model and the precise dimension of 
       brane depends on the boundary conditions along
       the other  sectors compatible with the GSO condition of the
       total system.  For instance, it is known that $\br_{\phi}$
       direction must  {\em always\/} obey the Neumann boundary condition
       to preserve the superconformal symmetry, as is mentioned by 
       several authors (for example, see \cite{eguchi}).}.  

We start with the affine supercurrents of $\widehat{SU}(2)_N$;
\begin{equation}
 \cJ^a (z,\theta) = \psi^a(z) + \theta J^a(z)~,
\end{equation} 
whose commutation relation is given by
\begin{equation}
 \cJ^a (z_1,\theta_1) \cJ^b (z_2,\theta_2) 
  \sim \frac{\frac{N}{2}\delta^{ab}}{z_{12}} + \frac{\theta_{12}}{z_{12}}
     i \ep^{ab}_{~~~c} \cJ^c (z_2,\theta_2) ~,
\end{equation}
 where $z_{12} = z_1 - z_2 - \theta_1 \theta_2$ and $\theta_{12} =
 \theta_1 -\theta_2$. 
In other words,
\begin{eqnarray}
 J^a (z_1) J^b (z_2) &\sim&  \frac{\frac{N}{2}\delta^{ab}}{(z_1 - z_2)^2} 
  + \frac{ i \ep^{ab}_{~~~c} J^c (z_2)}{z_1 -z_2}~, \nn
 J^a (z_1) \psi^b (z_2) &\sim& \psi^a (z_1) J^b (z_2) ~\sim~ 
   \frac{ i \ep^{ab}_{~~~c} \psi^c (z_2)}{z_1 -z_2}~, \nn
 \psi^a (z_1) \psi^b (z_2) &\sim& \frac{\frac{N}{2}\delta^{ab}}{z_1 - z_2}~.
 \end{eqnarray} 
It is also convenient to rewrite the ``total current'' $J^a$ by the ``bosonic
current'' $j^a$ and the ``fermionic current'' $j^a_f$;
\begin{equation}
 J^a (z) = j^a (z) + j^a _f (z) ~,~~ 
 j^a _f (z) = - \frac{i}{N} \ep^{a}_{~bc}  \psi^b (z) \psi^c (z) ~,
 \label{fcurrent}
\end{equation}  
which have the OPEs as follows
\begin{eqnarray}
 j_f ^a (z_1) j_f ^b (z_2) &\sim&  
  \frac{\delta^{ab}}{(z_1 - z_2)^2} 
  + \frac{ i \ep^{ab}_{~~~c} j_f ^c (z_2)}{z_1 -z_2} ~,\nn
 j^a (z_1) j^b (z_2) &\sim& 
  \frac{\frac{N-2}{2}\delta^{ab}}{(z_1 - z_2)^2} 
  + \frac{ i \ep^{ab}_{~~~c} j^c (z_2)}{z_1 -z_2} ~,\nn
 j^a (z_1) j_f ^b (z_2) &\sim& 0~.
 \end{eqnarray} 
In fact, $j^a(z)$ is identified with the current defined 
by the bosonic sector of super WZW model.

The $N=1$ superconformal algebra has a energy-momentum tensor and a
superconformal current. In the superfield formalism, we can combine
these currents as
\begin{eqnarray}
 \cT(z ,\theta) &=& \frac{1}{2} G(z) + \theta T(z) \nn 
   &=& \frac{1}{N} : D \cJ^a (z, \theta) \cJ^a (z, \theta) :
   + \frac{2i}{3N^2}  \ep_{abc} 
       \cJ^a (z, \theta) \cJ^b (z, \theta)\cJ^c (z, \theta)  ~, 
\label{supercurrent}
\end{eqnarray} 
where we set $D= \frac{\partial}{\partial \theta} + \theta
\frac{\partial}{\partial z}$. 
%
We here note that the zero-modes of the total currents $J^a$ commute with 
all the modes of  superconformal currents;
\begin{equation}
[G(z),J^a_0] = [T(z),J^a_0] =0 ~. 
\label{GJTJ}
\end{equation}


The supersymmetric extensions of the gluing conditions 
\eqn{glue1}, \eqn{glue2} are given by 
\begin{eqnarray}
 (J^a_n + \tJ^a_{-n}) \ket{B;\ep} &=& 0 ~,\nn
 (\psi^a_r +  i \ep \tpsi^a_{-r}) \ket{B;\ep} &=& 0~, \nn 
 (L_n - \tL_{-n}) \ket{B;\ep} &=& 0~, \nn
 (G_r - i \ep \tG_{-r}) \ket{B;\ep} &=& 0~, 
 \label{N=1}
\end{eqnarray}
where $\ep =\pm 1$ indicates the signature related to
the choice of NS or R sectors in the open string channel.

For the later convenience we introduce some notations of the currents at the
boundary just as in the bosonic case 
\begin{eqnarray}
 J^a_{\pm}(\sigma) &=& J^a(\sigma) \pm \tJ^a(\sigma) ~, \nn
 \psi^a_{\pm}(\sigma) &=&  \psi^a(\sigma) \pm   i \ep \tpsi^a(\sigma)~, \nn
 T_-(\sigma) &=& T(\sigma) - \tT(\sigma) ~, \nn
 G_-(\sigma) &=& G(\sigma) - i \ep \tG(\sigma) ~,
\end{eqnarray}
where we set
\begin{equation}
 \psi^a _{\pm} (\sigma) = \sum _n \psi^a _{\pm ,n} e^{- i n \sigma}
 ~,~~  \psi^a _{\pm ,n} = \psi^a _n \pm i \ep \tpsi^a _{-n}~.
\end{equation}

By using these notations the gluing conditions are expressed as
\begin{eqnarray}
J^a_+(\sigma) \ket{B;\ep} &=& 0 ~, \nn
\psi^a_+(\sigma) \ket{B;\ep} &=& 0 ~, \nn  
T_-(\sigma)\ket{B;\ep} &=& 0 ~, \nn
G_-(\sigma) \ket{B;\ep} &=& 0 ~.
\label{s glue}
\end{eqnarray}
However, as in the bosonic case,  it is really enough to only impose 
the conditions
\begin{equation}
J^a_+(\sigma) \ket{B;\ep} = 0,~~~ \psi^a_+(\sigma) \ket{B;\ep} = 0~,
\label{s glue 2}
\end{equation}
or equivalently,
\begin{equation}
j^a_+(\sigma) \ket{B;\ep} = 0,~~~ \psi^a_+(\sigma) \ket{B;\ep} = 0~,
\label{s glue 3}
\end{equation}
since the superconformal invariance at the boundary can be  derived 
from these conditions.


We also  define the currents 
\begin{equation}
 j^a _{f,\pm} (\sigma)= j^a _f (\sigma)\pm \tj^a_f (\sigma)~,
\end{equation}
where the fermionic currents $j^a _f$, $\tj^a_f$ are given by (\ref{fcurrent}) 
and we can rewrite them as 
\begin{eqnarray}
j^a _{f,+} (\sigma) &=& - \frac{i}{2N} \epsilon^{abc} 
 ( \psi^b_+ \psi^c_- + \psi^b_- \psi^c_+  ) ~, \nn
j^a _{f,-} (\sigma) &=& - \frac{i}{2N} \epsilon^{abc} 
 ( \psi^b_+ \psi^c_+ + \psi^b_- \psi^c_-  ) ~.
\end{eqnarray}

\subsection{Wilson Line in $SU(2)$ Super WZW Model}

We extend the discussion about the Wilson line in bosonic WZW model 
to the supersymmetric case. We shall start with the superfield
representation of boundary currents; 
\begin{equation}
\cJ^a_{\pm} (\sigma,\theta) \equiv \psi^a_{\pm}(\sigma)+\theta 
J^a_{\pm}(\sigma),~~~
\cT_-(\sigma, \theta) \equiv \frac{1}{2}G_-(\sigma)+\theta T_-(\sigma)~ .
\label{b super currents}
\end{equation}
The gluing conditions \eqn{s glue} are  now given by 
\begin{eqnarray}
 \cJ^a _+ (\sigma,\theta) \ket{B;\ep} &=& 0    ~, \nn
 \cT_-(\sigma,\theta) \ket{B;\ep} &=& 0~.
\end{eqnarray}


A natural extension of the bosonic Wilson line \eqn{path integral}
to the supersymmetric one is given by 
\begin{equation}
 W(\{M^a\})\equiv
 \int \cD \cG \exp \left[  i \int ^{2 \pi} _0 d \sigma \int d \theta\,
   \bra{\cG} \left(D - \frac{1}{N} \cJ_-^aM^a\right) \ket{\cG}\right]~,
\label{superCP}
\end{equation}
where the supercoordinate $\cG(\sigma,\theta)$ is defined as 
\begin{equation}
 \cG (\sigma , \theta) = \exp(i \theta \eta^a R_L(T^a)) g(\sigma)~.
\end{equation}
($\eta^a(\sigma)$ are Grassmann coordinates.)
The path-integral measure is defined 
in the standard way, $\cD \cG = \cD g \cD \eta$. Precisely speaking, 
we must define it so that it does not include the zero-modes of 
the fermionic coordinates $\eta^a$ to obtain non-vanishing integrals.  
We also used the symbol of the superderivative defined by
\begin{equation}
 D = - \frac{\partial}{\partial \theta} 
  + \theta i \frac{\partial}{\partial \sigma}~.
\end{equation}

We now focus on  the two cases of CP matrices just as in the bosonic 
case; (1) the commutative CP matrices $\lb M^a,~ M^b\rb=0$ 
and (2) the CP matrices of fuzzy sphere 
$\lb M^a,~M^b \rb =i\ep^{ab}_{~~c}M^c$.

For the first case we assume for simplicity
that $M^1=M^2=0$ and $M^3$ is a diagonal matrix. 
By this assumption we can readily carry out the $\eta$-integral,
since this is nothing but a Gaussian integral, and obtain
\begin{equation}
W(\{M^a\}) = \tr \left(\exp\left(-\frac{2\pi i}{N}J^3_{-,0}M^3\right)\right)~.
\label{s commutative}
\end{equation} 
This Wilson line actually preserves the world-sheet superconformal
invariance because of the commutation relations \eqn{GJTJ}.


For the second case of fuzzy sphere
we can again integrate out the fermionic coordinate 
$\eta(\sigma)$ explicitly. However, we have to be more careful 
to evaluate the Gaussian integral. 
We again assume that $M^a=R_L(T^a)$ and make use of the abbreviated
notations $\cJ_-\equiv \cJ_-^a M^a$, $\psi_-\equiv \psi_-^aM^a$, etc.
in the following discussion.
We first note that 
\begin{eqnarray}
 \lefteqn{\int \cD \cG \exp \left[  i\int ^{2 \pi} _0 d \sigma 
  \int d \theta \, \bra{\cG} \left(D - \frac{1}{N} \cJ_-\right) 
   \ket{\cG}\right]} \nn
 &=& \int \cD g \cD \eta \exp \left[  - \int ^{2 \pi} _0 d \sigma 
    \bra{g} \left\{\left(  \frac{d}{d \sigma} + \frac{i}{N}J_-\right)
     + i  \eta \eta  + \frac{1}{N}\left(\eta\psi_- + \psi_-\eta\right) \right\}
    \ket{g} \right]  ~.
\label{superCP2}
\end{eqnarray}
Thus we can apply the Gaussian  integral  
\begin{equation}
  \int \cD \eta \exp \left[  -\int ^{2 \pi} _0 d \sigma
    \bra{g}  i  \eta \eta \ket{g} \right] = C ~,
\end{equation}
where $C$ is some constant which is independent of $g$ 
because the path-integral measure  is invariant under the
transformation 
$\eta \to U(\sigma)\, \eta\, U(\sigma)^{-1}$.
Since the path integral (\ref{superCP2}) has the linear term
of $\eta$, we obtain
\begin{eqnarray}
 \lefteqn{ \int \cD \eta \exp \left[  - \int ^{2 \pi} _0 d \sigma \, 
    \bra{g}  \left(i  \eta \eta  + \frac{1}{N}
    \left(\eta \psi _- + \psi_-\eta \right)  \right)
    \ket{g} \right] }\nn
 &=&  \int \cD \eta \exp \left[  - \int ^{2 \pi} _0 d \sigma \,
    \left\{ \bra{g}  i (\eta  - \frac{i}{N} \psi _- )^2  \ket{g} 
    + \bra{g}  \frac{i}{N^2} {\psi _-} ^2  \ket{g}\right\} 
   \right]~,
\end{eqnarray}
and can apply the Gaussian  integral to the first term of the second line.   
Naively  it seems that the integral again gives merely the same constant
$C$ by means of the simple change of variable 
$\dsp \eta' = \eta - \frac{i}{N} \psi _-$.   But this is {\em not\/}
correct, since $\psi^a_-$ is not a c-number.
We have to carefully  evaluate the $g$ dependence of the path integral
\begin{equation}
 F(g) = \int \cD \eta \exp \left[  - \int ^{2 \pi} _0 d \sigma 
    \bra{g}  i \left(\eta  - \frac{i}{N} \psi _- \right)^2  \ket{g} 
    \right]~.
\end{equation}
First we point out that $F(g)$ is obviously a c-number, which does not
include the operator $\psi_-$. 
Replacing $g(\sigma)$ by $U(\sigma)g(\sigma)$, we obtain  
\begin{eqnarray}
 F(Ug) &=& \int \cD \eta \exp \left[  - \int ^{2 \pi} _0 d \sigma 
    \bra{g} U^{-1} i (\eta  -  \frac{i}{N} \psi _- )^2
    U \ket{g} \right] \nn
 &=&  \int  \cD \eta ' \exp \left[  -\int ^{2 \pi} _0 d \sigma 
    \bra{g}  i \left\{   \eta^{'2} - \frac{i}{N}
    \left(\eta' U^{-1}\psi_- U + U^{-1}\psi_- U \eta'  \right) 
\right. \right. \nn
 && \hspace{7cm} \left. \left.
    -\frac{1}{N^2} U^{-1}\psi^{2}_- U \right\}  \ket{g} 
   \right]~,
\end{eqnarray}
where we changed the integral 
variable $\eta  \to \eta' \equiv U^{-1}\, \eta \,  U$.
We next make use of  the following equalities  
\begin{eqnarray}
 U^{-1}(\sigma) \psi_- U(\sigma) &=& 
     \hat{U}^{-1} \psi_- \hat{U} ~, \\
 U^{-1}(\sigma) \frac{1}{N}{\psi_-}^2 U (\sigma)&=&
  \hat{U}^{-1} \frac{1}{N}{\psi_-}^2\hat{U}
  - 2i U(\sigma)^{-1} \frac{d}{d \sigma} U (\sigma)~.
\label{anomaly}
\end{eqnarray}
where we set $\dsp U(\sigma) = \exp (- 2 \pi i u^a (\sigma) M^a)$ and 
$\dsp \hat{U} = \exp 
\left( i \oint d \sigma  u^a (\sigma) j^a _{f,+}(\sigma) \right)$.
The second term  in the right hand side of 
the equality \eqn{anomaly} is originating from 
the anomalous contribution when evaluating it on the boundary state.
In fact, we point out that 
\begin{equation}
\frac{1}{N}\psi^2_-(\sigma)\ket{B;\ep} \equiv 
- j_{f,-}^a(\sigma)M^a \ket{B;\ep} ~, 
\end{equation}  
and 
\begin{equation}
\hat{U}^{-1}j^a_{f,-}(\sigma)\hat{U}
= U(\sigma)^{-1}j^a_{f,-}(\sigma)U(\sigma) 
- 2iU(\sigma)^{-1} \frac{d}{d\sigma} U(\sigma) ~.
\end{equation} 
$F(Ug)$ is then evaluated as 
\begin{eqnarray}
  \lefteqn{F(Ug) } \nn
  &=& \hat{U}^{-1}
   \int \cD \eta' \exp \left[  - \int ^{2 \pi} _0 d \sigma 
    \bra{g} i  \left(\eta '  -\frac{i}{N} \psi _- \right)^2
     \ket{g} 
    - \frac{2}{N} \bra{g} U(\sigma)^{-1} \frac{d}{d \sigma} U (\sigma) 
   \ket{g}\right] \hat{U}  \nn
   &=& F(g) \exp  \left[  - \int ^{2 \pi} _0 d \sigma 
   \left\{ - \frac{2}{N} \bra{g} U(\sigma)^{-1} \frac{d}{d \sigma} U (\sigma) 
   \ket{g}\right\}\right]~.
\end{eqnarray} 
We can thus obtain
\begin{equation}
 F(g) = C '  \exp  \left[  -  \int ^{2 \pi} _0 d \sigma 
    \left\{- \frac{2}{N} \bra{g}  \frac{d}{d \sigma}  
   \ket{g} \right\}\right] ~,
\end{equation}
where  $C '$ is some constant independent of $g$.

In this way we can obtain from \eqn{superCP2} 
\begin{eqnarray}
  &~& \int \cD g \exp \left[  - \int ^{2 \pi} _0 d \sigma 
    \bra{g} \left( \left(1-\frac{2}{N}\right) 
  \frac{d}{d \sigma} + \frac{i}{N}J_-\right)
     +\frac{i}{N^2}{\psi_-}^2
    \ket{g} \right] \nn
 &~&~~~~~=~ \int \cD g \exp \left[- \int ^{2 \pi} _0 d \sigma 
    \bra{g} \left( \left(1-\frac{2}{N}\right) 
    \frac{d}{d \sigma} + \frac{i}{N}j_-\right)
    \ket{g} \right]  \nn
 & ~& ~~~~~=~\int \cD g \exp \left[-  \frac{N-2}{N} \int ^{2 \pi} _0 d \sigma 
    \bra{g} \left(\frac{d}{d \sigma} + \frac{i}{N-2}j_-\right)
    \ket{g} \right]~. 
\end{eqnarray}
On the second line  we used
 $J_- + \frac{1}{N} \psi^2_- =
 j_- - \frac{1}{N} \psi^2_+$ and
neglect the term 
 $ - \frac{1}{N} \psi^2_+$ because of the gluing condition 
 $\psi^a _+  (\sigma) \ket{B;\ep} = 0$. 
As we already pointed out, the factor $\dsp \frac{N-2}{N}$ is irrelevant
and we finally obtain for the CP matrices of fuzzy sphere
\begin{equation}
W(\{M^a\}) = \tr \left( P \exp \left(-\frac{i}{N-2}
\int_0^{2\pi} d\sigma j^a_-(\sigma)M^a\right)\right) ~.
\label{superCP3}
\end{equation}

This form is the same as the Wilson line in the bosonic case.
It is quite important that the final expression 
\eqn{superCP3} contains only the {\em bosonic\/} currents $j^a_-$
in contrast to the case of commutative CP matrices \eqn{s commutative}, 
in which the {\em total\/} currents $J^a_-$ appear instead of $j^a_-$.
Because we already know that the Wilson line \eqn{superCP3} 
commutes with all the currents $j^a_+(\sigma)$ 
(and also $J^a_+(\sigma)$), it is straightforward to show 
that this Wilson line preserves the superconformal invariance.
(Recall that the bosonic currents $j^a(z)$ have the level $N-2$.)
It is also not difficult to show  the same property 
directly  from the expression \eqn{superCP}
in the similar argument as in the bosonic case. 
In fact, we can prove the supersymmetric extension of the equality 
\eqn{DD};
\begin{equation}
\hat{U}\left(D-\frac{1}{N}\cJ_-(\sigma, \theta)\right)\hat{U}^{-1}
=U(\sigma,\theta)\left(D-\frac{1}{N}\cJ_-(\sigma, \theta)\right)
U(\sigma,\theta)^{-1}~,
\label{sDD}
\end{equation}
where we set 
\begin{equation}
\begin{array}{l}
\dsp  \hat{U}\equiv\exp\left(i\int_0^{2\pi}d\sigma\int d\theta\, 
\bu^a(\sigma,\theta)\cJ_+^a(\sigma,\theta) \right),~~~
U(\sigma,\theta)\equiv \exp\left(-2\pi i \bu^a(\sigma,\theta)M^a\right)~.\\
\hspace{2cm}
(\bu^a(\sigma,\theta) = u^a(\sigma)+\theta \ep^a(\sigma);
 ~~~\mbox{$\ep^a$ are Grassmannian.})  
\end{array}
\end{equation}
Hence we can likewise prove that the supercurrents 
$\cJ_+^a(\sigma,\theta)$ commute with the Wilson line 
of fuzzy sphere.

\subsection{Boundary State Analysis for the D-branes in the  NS5 Background
 and Space-time  SUSY}

Now let us study the D-branes in the  NS5 background. 
We  shall consider the near horizon geometry of $(N-2)$ NS5-branes 
described by the CHS $\sigma$-model \cite{chs};
\begin{equation}
\R^{5,1}\times \R_\phi \times SU(2)~ ,
\end{equation}
where $\R_\phi$ denotes the radial direction and is described by the Liouville
theory. The background charge of Liouville field is given by 
$Q=\sqrt{2\over N}$. $SU(2)$ stands for  the $SU(2)$ WZW model (the level
of bosonic currents is equal to $N-2$). 
There are also 8 (physical) free fermions, which make the $SU(2)$ sector
raised to the super WZW model with the shifted level $N$.
Since the bosonic coordinates along $\R^{5,1}\times \br_{\phi}$ 
are not important for our discussion, we will neglect these parts. 
Nevertheless we need take account of 
the fermionic coordinates of these directions 
to impose the GSO condition properly. 
We shall hence deal with  the system of $SU(2)$ WZW model and 8 
free fermions. 
We here only work on the structure of $N=1$ world-sheet SUSY, 
although it is known that the world-sheet SUSY 
of the CHS $\sigma$-model can be extended to
$N=4$ \footnote
   {We must incorporate the $\br_{\phi}$-direction for the
    extended SUSY. (See appendix B.)}.
We summarize several features of our system as 
superconformal theory with the extended SUSY in appendix B.

We begin our analysis by introducing
the Ishibashi states for the system of interest;
\begin{equation}
|\ell,s\rangle_I ~,\quad\quad \ell =0,1,\ldots ,N-2 ~,~~~s\in \Z _4 ~,
\end{equation}
where $s$ labels the integral representations of $\widehat{SO}(2n)_1$
($n=4$) in our setup 
($s=0$ for basic, $s=2$ for vector, $s=1$ for spinor and $s=3$ for
cospinor representations, respectively).
These  states satisfy the following boundary conditions
\begin{eqnarray}
&&(J_n^a+ \widetilde J_{-n}^a )|\ell,s\rangle_I=0~,\n\\
&&\psi^a_r|\ell,s\rangle_I=-i\tpsi^a_{-r}|\ell,s+2\rangle_I~,\n\\
&&(L_n-\widetilde L_{-n} )|\ell,s\rangle_I=0~,\n\\
&&G_r|\ell,s\rangle_I=i\widetilde G_{-r}|\ell,s+2\rangle_I~.
\end{eqnarray}

The ``inner products'' between Ishibashi states are given by 
\begin{equation}
{}_I\langle\ell,s |\tilde q^{H^{(c)}}|\ell',s'\rangle_I=
\delta_{\ell,\ell'}\delta_{s,s'}(-1)^s
\chi_{\ell}^{(N-2)}(\tilde q)\chi_s^{SO(8)}(\tilde q)~,
\end{equation}
where $\chi_s^{SO(8)}$ is the character of $\widehat{SO}(8)_1$.
In general, 
the character of $\widehat{SO}(2n)_1$ is given by
\begin{eqnarray}
\chi_0^{SO(2n)} +\chi_2^{SO(2n)} =\left(\theta_3\over \eta
\right)^{n}, &\quad&
\chi_0^{SO(2n)} -\chi_2^{SO(2n)} =\left(\theta_4\over \eta 
\right)^{n},\n\\
\chi_1^{SO(2n)} +\chi_3^{SO(2n)}=\left(\theta_2\over \eta 
\right)^{n}, &\quad&
\chi_1^{SO(2n)} -\chi_3^{SO(2n)} =\left(-i\theta_1\over \eta 
\right)^{n}.
\end{eqnarray}
Then, we can construct the Cardy states 
\begin{equation}
|L,S\rangle_C =\sum_{\ell=0}^{N-2}\sum_{s\in \bsz_4}\, 
{S_{L\ell}\over \sqrt{S_{0\ell}}}
{e^{{-i\pi}{Ss\over 2}}\over \sqrt 2}|\ell,s\rangle_I~,
\end{equation}
where $S$ labels spin structure.
The cylinder amplitude between these Cardy states is given by
\begin{equation}
_C\langle L',S'| \tilde q^{H^{(c)}}|L,S\rangle_C
=\sum_{\ell,s}N_{L,L'}^{\ell}\chi_{\ell}^{(N-2)}(q)
\chi_{s}^{SO(8)}(q)\delta^{(4)}_{S-S'+s+2,0}~,
\end{equation}
where $\delta^{(4)}_{a,0}$ defines $a=0$ mod 4.

It is also convenient to introduce the projections of 
the Cardy states to the  NSNS-sector 
$|L;\ep\rangle_C^{(NS)}$ ($\ep=\pm1$) and to the 
RR-sector $|L;\ep\rangle_C^{(R)}$;
\begin{eqnarray}
|L;\ep\rangle_C^{(NS)}&=&
    |L,S\rangle_C+|L,S+2\rangle_C
~,\n\\
|L;\ep\rangle_C^{(R)}&=&
  |L,S\rangle_C-|L,S+2\rangle_C ~,
\label{s Cardy}
\end{eqnarray}
where $\ep=+1$ corresponds to $S=0$ and $\ep=-1$ corresponds to $S=1$.
These boundary states satisfy the boundary condition (\ref{N=1})
and  especially preserve the $N=1$ world-sheet SUSY.

We must suitably take account of the GSO projection 
$\frac{1-(-1)^F}{2}$ in the underlying string theory, 
where $F$ denotes the world-sheet fermion number in the left mover.
Since the action of operator $(-1)^F$ to the Cardy state is given by
\begin{equation}
(-1)^F|L,S\rangle_C=|L,S-1\rangle_C~,
\end{equation}
the GSO invariant combinations in NSNS-sector and RR-sector are 
\begin{eqnarray}
|B;L\rangle^{(NS)}
&=&{1\over \sqrt{2}}\left(|L;+1\rangle_C^{(NS)}
-|L;-1\rangle_C^{(NS)}\right)~,\n\\
|B;L\rangle^{(R)}
&=&{1\over \sqrt{2}}\left(|L;+1\rangle_C^{(R)}
+|L;-1\rangle_C^{(R)}\right)~.
\end{eqnarray}
Then the supersymmetric boundary state representing a D-brane is 
defined  by
\begin{equation}
|B;L\rangle={1\over \sqrt{2}}\left(|B;L\rangle^{(NS)}+|B;L\rangle^{(R)}
\right)~.
\end{equation}
The boundary state describing the anti-brane is likewise given by
\begin{equation}
|\bar{B};L\rangle={1\over \sqrt{2}}\left(|B;L\rangle^{(NS)}-|B;L\rangle^{(R)}
\right)~.
\end{equation}

We  calculate the cylinder amplitude between the branes. If the brane
configuration preserves the  space-time SUSY, the amplitude should
vanish. The amplitude between the branes labeled by $L$ and $L'$ becomes
\begin{equation}
\langle B;L | \tilde q^{H^{(c)}} | B;L' \rangle  =\sum_{\ell}N_{L,L'}^{\ell}
\chi_{\ell}^{(N-2)}
\left[\left({\theta_3\o\eta}\right)^4
-\left({\theta_4\o\eta}\right)^4
-\left({\theta_2\o\eta}\right)^4
\right](q)=0~,
\label{M=0}
\end{equation}
which implies the existence of space-time SUSY as pointed out.
On the other hand, the amplitude between the brane and
anti-brane becomes
\begin{equation}
\langle {\bar B};L | \tilde q^{H^{(c)}} 
| B;L' \rangle =\sum_{\ell}N_{L,L'}^{\ell}
\chi_{\ell}^{(N-2)}
\left[\left({\theta_3\o\eta}\right)^4
+\left({\theta_4\o\eta}\right)^4
-\left({\theta_2\o\eta}\right)^4
\right](q)\neq 0 ~,
\end{equation}
as expected, since this  configuration 
does not preserve any supersymmetry.

Now our main purpose is to study the aspects of boundary states
with the insertion of  Wilson line previously considered. 
The Cardy state
$\ket{B;L=0}$ is the single ``D0-brane state'', and 
the various configurations of multiple D0-branes are realized as 
\begin{equation}
\ket{B;\{M^a\}} = W(\{M^a\})\ket{B;L=0}~,
\label{s boundary with WL}
\end{equation}
where $W(\{M^a\})$ is the Wilson line operator defined by \eqn{superCP}.

Suppose $M^a$ are the $(L+1)\times (L+1)$ matrices of the fuzzy sphere,
then the boundary state $\ket{B;\{M^a\}}$ satisfies the gluing
conditions \eqn{s glue}, as we already discussed. 
It is also  clear that $\ket{B;\{M^a\}}$ preserves the half of space-time
SUSY, since the expression \eqn{superCP3} does not contain the fermionic
degrees of freedom and thus it obviously commutes with any spin fields.
Moreover we can obtain, under the large $N$ limit,
\begin{equation}
\ket{B;\{M^a\}} \approx \ket{B;L}~,
\label{s formation}
\end{equation}
just as in the bosonic case.
The right hand side is the Cardy state 
corresponding  to the D2-brane  
wrapped on the $(L+1)$-th conjugacy class of $SU(2)$.

Next we consider the commutative CP matrices and again we assume that 
$M^1=M^2=0$ and $M^3= \mbox{diag}(a_1,a_2,\ldots, a_{L+1})$.
The boundary states $\ket{B;\{M^a\}}$ now becomes 
\begin{equation}
\ket{B;\{M^a\}} =  
\tr \left(\exp\left(-\frac{2\pi i}{N}J^3_{-,0}M^3\right)\right) \ket{B; 0}
= \sum_{i=1}^{L+1}\ket{B;L=0, a_i}~,
\label{s bs comm}
\end{equation}
where we set 
\begin{equation}
\ket{B;L,a} \equiv e^{-4 \pi i {J_0^3\over N} a} \ket{B;L}~.
\label{bla}
\end{equation}

We now observe that the boundary states \eqn{s bs comm}
cannot preserve any space-time SUSY except for the special case
$a_1=a_2=\cdots=a_{L+1}$. In fact, let $Q_{\ep_1 \cdots \ep_4}$,
$\tilde{Q}_{\bar{\ep}_1 \cdots \bar{\ep}_4}$ 
($\ep_i,\bar{\ep}_i=\pm 1$) be the space-time SUSY charges
defined by the four real bosons  $H_1,\, \ldots, H_4$ bosonizing 
the transverse fermions. Especially, we take the convention such 
that 
\begin{equation}
i\partial H_1 = i\psi^1\psi^2 ~,
\end{equation} 
and its counter part of the right mover.
Assume that $\ket{B;L=0}$ preserves the following SUSY with some
coefficients $\La_{\ep_1\cdots}^{~\bar{\ep}_1\cdots}$;
\begin{equation}
\left(Q_{\ep_1\cdots} + \La_{\ep_1\cdots}^{~\bar{\ep}_1\cdots}
\tilde{Q}_{\bar{\ep}_1\cdots}\right)\ket{B;L=0}=0~.
\end{equation}
Since the total current $J^3$ has a non-trivial  commutation relations
with $H_1$, we can find that 
\begin{equation}
Q_{\ep_1\cdots} e^{-\frac{4\pi i}{N} J^3_0\,a}
   = e^{-\frac{2\pi i \ep_1 a}{N}}\,
e^{-\frac{4\pi i}{N}J^3_0\,a} Q_{\ep_1\cdots}  ~.  
\end{equation}  
Hence the boundary state $\ket{B;L=0,a_i}$ preserves the space-time SUSY
\begin{equation}
\left(Q_{\ep_1\cdots} +  \La_{\ep_1\cdots}^{~\bar{\ep}_1\cdots}
 e^{-\frac{2\pi i \ep_1 a_i}{N}}
\tilde{Q}_{\bar{\ep}_1\cdots}\right) \ket{B;0,a_i} =0~,
\end{equation} 
but their summation \eqn{s bs comm} does {\em not\/} except for the  case 
$a_1=a_2=\ldots=a_{L+1}$.

We  also examine whether or not the brane configuration preserves 
the space-time SUSY by another way.
After some straightforward calculations 
we find out the following cylinder amplitudes
\begin{eqnarray}
&&\bra{B;L',a'}\tilde q^{H^{(c)}}\ket{B;L,a} \nn
&&= \sum_{\ell,m}N_{L,L'}^\ell
\left[
\left({\theta_3\over \eta}\right)^3\C_{\ell,m}^{(NS)}
-\left({\theta_4\over \eta}\right)^3{\widetilde \C}_{\ell,m}^{(NS)}
-\left({\theta_2\over \eta}\right)^3\C_{\ell,m}^{(R)}
\right]{\theta_{m+2(a'-a),N}\over \eta}(q)~,
\label{cyl amp}
\end{eqnarray}
where $\C_{\ell,m}^{(*)}(q)$ is the character of $N=2$ minimal model
of level $N-2$ (see appendix C). In this calculation  the next 
``Gepner model like'' reinterpretation  of CHS $\sigma$-model 
is essential \cite{ov,gkp,gepner};
\begin{equation}
 \br_{\phi} \times \mbox{super} ~ \widehat{SU}(2)_N 
  \cong \frac{(\br_{\phi}\times S^1) \times \left(SU(2)_N/U(1)\right)}
 {\bz_N}~,
\end{equation}
where $\br_{\phi}\times S^1$ denotes the $N=2$ Liouville theory 
with $\dsp \hat{c}\equiv 1+ Q^2 = \frac{N+2}{N}$ and 
$SU(2)_N/U(1)$ denotes the Kazama-Suzuki model \cite{ks} 
for $SU(2)/U(1)$,
which is one of the concise realization of the $N=2$ minimal 
model of level $N-2$ ($\dsp \hat{c}=\frac{N-2}{N}$). Note that 
$J^3_0$ commutes with all the generators of this Kazama-Suzuki model
and the appearance of the minimal characters $\C_{\ell,m}^{(*)}(q)$ 
is due to this fact.
Especially, we here only need the $L=L'=0$ sector of \eqn{cyl amp};
\begin{eqnarray}
&&\bra{B;0,a'}\tilde q^{H^{(c)}}\ket{B;0,a} \nn
&&= \sum_{m}
\left[
\left({\theta_3\over \eta}\right)^3\C_{0,m}^{(NS)}
-\left({\theta_4\over \eta}\right)^3{\widetilde \C}_{0,m}^{(NS)}
-\left({\theta_2\over \eta}\right)^3\C_{0,m}^{(R)}
\right]{\theta_{m+2(a'-a),N}\over \eta}(q)~.
\label{cyl amp 2}
\end{eqnarray}

When $a=a'$ holds, the above amplitude \eqn{cyl amp} 
reduces to the previous one (\ref{M=0}) because of the 
``branching relation'' \eqn{branching relation} and vanishes,
which reflects the existence of space-time SUSY.
On the other hand, when $a \neq a'$, the amplitude \eqn{cyl amp} 
(and, of course, \eqn{cyl amp 2}) dose not vanish. 
This implies that the configuration of branes 
characterized by the different $a$'s  does not preserve any 
space-time SUSY. It is also easy to show that 
the cylinder amplitude defined with respect to 
the boundary state \eqn{s bs comm}
does not vanish except for  the case of $a_1=\ldots=a_{L+1}$,
which confirms our above expectation.

Since $a_1,\, \ldots,\,a_{L+1}$ correspond to the positions of 
$(L+1)$ D0-branes on $S^3$, the above result means that  
{\em only\/} the stack of D0-branes should be BPS among the commutative 
configurations. This fact  contrasts to the aspects in  the 
flat backgrounds, in which we can freely distribute multiple D0-branes
without breaking the space-time SUSY. Arbitrary commutative
configurations of D0-branes in the flat backgrounds are marginally stable. 
In our case of $S^3$  we can expect that the non-BPS 
configuration \eqn{s bs comm} (with, say, $a_1\neq a_2$) 
should be unstable. In fact, we can find out 
the tachyonic excitations by studying  the spectrum 
in the open string channel.
For example, observing the $q$-expansion of the cylinder amplitude
\eqn{cyl amp 2} (when setting $a=a_1$, $a'=a_2$), we can show that 
the lowest mass in NS-sector is evaluated as 
\begin{equation}
({\rm mass})^2 =  {(\frac{1}{2}m+\Delta a)^2 \over N}
-\frac{1}{4}{m^2\over N}+ \frac{Q^2}{8}
={\Delta a(\Delta a+m)\over N} + \frac{1}{4N}~,
\label{tachyon mass}
\end{equation}
where $\Delta a = a_1-a_2$ and $|\Delta a|\leq N/2$. 
The term $Q^2/8\equiv 1/(4N)$ is the contribution from 
the mass gap in the Liouville sector $\br_{\phi}$, which  
we neglected in the above argument.  This term is not important 
under the large $N$ approximation. 
Therefore, if $\Delta a \neq 0$ holds, there always exists 
a particular $m\in \Z_{2N}$ which generates a negative mass squared. 
In other words, we always have an open string tachyon in the cases of  
$\Delta a \neq 0$, which makes the system unstable.
We will next discuss how this instability is related to the formation 
of spherical D2-brane.


\subsection{Tachyon Condensation and Formation of Spherical Brane}

As is discussed in the flat background in many literatures \cite{tachyon}, 
we can often expect that  some stable BPS configurations of branes arise 
after the open string tachyons condensate. 
It is quite interesting to discuss the similar phenomenon 
in our case of the NS5 background.
For the simplicity we shall take a simple example $L=1$,
namely, the case of two D0-branes on $S^3$. We start with 
the $2\times 2$ CP matrices $M^1=M^2=0,$ $M^3=\sigma^3 /2$,
which means that the Wilson line \eqn{superCP} contains 
the super affine currents as the form 
\begin{equation}
\cJ^a_-M^a = \frac{1}{2}\left(
\begin{array}{cc}
 \cJ^3_-(\sigma)& 0 \\
  0 & -\cJ^3_-(\sigma)
\end{array}
\right)~.
\label{CP non-BPS}
\end{equation}
As we previously observed, this is a non-BPS configuration and 
we have the tachyonic modes. 
The tachyon fields should arise as the non-diagonal elements of CP
matrices, just like  the $D-\bar{D}$ system in the flat background.
Therefore it is reasonable to consider the following deformation of  
\eqn{CP non-BPS};
\begin{equation}
\cJ^a_-M^a= \frac{1}{2}
\left(
\begin{array}{cc}
 \cJ^3_-(\sigma)& \cJ^-_-(\sigma) T \\
 \cJ^+_-(\sigma) \bar{T}  & -\cJ^3_-(\sigma)
\end{array}
\right)~,
\label{CP non-BPS 2}
\end{equation}
where we introduced the ``tachyon field'' $T$, $\bar{T}(\equiv
T^{\dag})$. (We here only consider the constant tachyon fields.)
At first glance this deformation $\eqn{CP non-BPS 2}$ seems 
to be marginal, but our observation about the mass spectrum of 
open string implies that it should be precisely {\em marginally relevant\/}.
Therefore the tachyon fields $T$, $\bar{T}$ glow along the trajectory
of RG flow starting from the point \eqn{CP non-BPS}. Then, can we have
the fixed point at which the tachyon fields condensate? 
It is a difficult problem to make a complete answer, since we need 
to solve the dynamics away from the conformal point.
Nevertheless, we can expect the next natural solution for the fixed point 
\begin{equation}
T = e^{i\alpha}, ~~~\bar{T}=e^{-i\alpha}~~~(\alpha\in \br)~.
\label{condensation}
\end{equation}
This is no other than the fuzzy sphere configuration! 
As we already observed, the CP matrices of fuzzy sphere  
correspond to a BPS bound state, and at least under the large $N$
approximation it is identified with the Cardy state (with $L=1$). 
Among other things, it is easy to show that the open string spectrum 
appearing in the cylinder amplitude like \eqn{M=0} has no tachyonic 
excitations. This means that the system is stable and resides at 
a fixed point of the boundary  renormalization group.

Our claim here is summarized as follows; 
\begin{equation}
\ket{B;\{M^a\}}\equiv
\ket{B;L=0,a=\frac{1}{2}}
+\ket{B;L=0,a=-\frac{1}{2}} ~ \stackrel{\msc{tachyon condensation}}
                           {\Longrightarrow} ~
\ket{B;L=1} ~.
\label{tachyon condensation}
\end{equation}

~

To close this subsection let us make several comments;

\noindent
1. It may be natural to assume that the central charge, 
which is directly  calculated from the RR-part of the boundary state 
\cite{ooy,divecchia,bds,eguchi}, should not change through the 
tachyon condensation. 
In fact, we can easily check that the both
sides of  \eqn{tachyon condensation} have the equal central charges.
More generically, the central charge of $\ket{B;\{M^a\}}$
with $M^1=M^2=0$ $M^3= \mbox{diag}(L/2, L/2-1, \ldots ,-L/2)$ is 
computed, up to some factors of no interest, as follows\footnote
     {Strictly speaking, we must turn on the Liouville potential term
      in the $N=2$ Liouville sector in order to obtain the non-zero central
      charges. See \cite{eguchi}.};
\begin{equation}
\sin\left(\frac{\pi}{N}\right) \times 
\left(e^{i\pi \frac{L}{N}}+e^{i\pi \frac{L-2}{N}}+\cdots 
+e^{-i\pi \frac{L}{N}}\right) \equiv \sin \left(\pi \frac{L+1}{N}\right)~,
\label{central charge}
\end{equation} 
which is indeed equal to the central charge of the Cardy state $\ket{B;L}$.

~

\noindent
2. The brane mass (or tension) can be readily read off from the NSNS
part of the boundary state \cite{divecchia,hkms}. 
The left hand side in \eqn{tachyon condensation} has the brane mass 
\begin{equation}
 \mbox{mass} \sim |\sin\left(\frac{\pi}{N}\right)|
  + |\sin\left(\frac{\pi}{N}\right)|~,
\label{brane mass 1}
\end{equation}
and the right hand side has 
\begin{equation}
\mbox{mass} \sim |\sin\left(\frac{2\pi}{N}\right)|
\label{brane mass 2}~.
\end{equation}
Clearly \eqn{brane mass 1} is greater than \eqn{brane mass 2} 
because of the triangular inequality. This feature reflects directly
the fact that the left hand side expresses the non-BPS branes, while 
the right hand side corresponds to the BPS saturated configuration.
It is also consistent with the $g$-theorem about the boundary RG flow 
\cite{g-theorem}.

~

\noindent
3. It is easy to extend \eqn{tachyon condensation} to more general
cases. Suppose that we start with the $(L+1)\times (L+1)$
CP matrices $M^1=M^2=0$, and $M^3$ is a diagonal matrix. 
Let us further assume that there exists an $(L+1)$-dimensional
(not necessarily irreducible) representation $R$ of $SU(2)$ such that
$R(T^3)=M^3$. In this situation, when we have the decomposition
\begin{equation}
R \cong R_{L_1} \oplus \cdots \oplus R_{L_r}~, ~~~ 
(\sum_{i=1}^r\,(L_i+1) = L+1 )~,
\end{equation}    
our claim should be addressed as follows;
\begin{equation}
\ket{B;\{M^a\}} ~\stackrel{\msc{tachyon condensation}}
                           {\Longrightarrow} ~
\sum_{i=1}^r \ket{B;L_i} ~.
\label{tachyon condensation 2}
\end{equation} 
It is not hard to show that the both sides of
\eqn{tachyon condensation 2} have the equal central charges 
and the total mass of left hand side is greater than 
that of the right hand side, which is decomposed to  the  $r$ pieces of 
Cardy states and describes  (marginally stable) BPS bound states.  
One might feel that the D0-brane configurations we are treating 
are rather limited, since the eigen-values of $M^3$ are now assumed 
to only take some discrete values. However, since our discussion here 
is based on the large $N$ approximation, one can expect that
the sufficiently dense distributions of D0-branes are realized in
this argument. (Recall the definition \eqn{bla}.)

~

\noindent
4. In the T-dualized framework, our NS5 background can be reinterpreted 
as the ALE space (with the vanishing $B$-field) of $A_{N-1}$ type
singularity \cite{ov,ghm}.  
In the picture of ALE space the boundary states of ``D0-branes'' 
$\ket{L=0, a= M/2}$ ($M= N, N-2, \ldots$) correspond to the ``primitive
vanishing cycles'' which are in one to one correspondence with the simple 
roots of $A_{N-1}$, and the ``D2-brane states'' $\ket{B;L}$ 
($L\neq 0, N-2$) correspond to the supersymmetric cycles 
(special Lagrangian submanifolds) homologous to the non-trivial sums of
the primitive vanishing cycles \cite{lerche,eguchi}. 
It is an interesting point that the fuzzy sphere configurations 
in the NS5 background is equivalent to the special Lagrangian 
configurations in the ALE side. The BPS saturation in the former 
seems to be due to a stringy effect, 
``fuzziness of the space-time coordinates'', while that of the latter
is based on the classical geometry with no quantum corrections.

~

\section{Summary and Discussions}

In this paper we especially studied the BPS bound states
of multiple D0-branes realized as the spherical D2-branes,
which was proposed in \cite{alek1,alek2}, from the view points 
of boundary states. We realized the configurations of D0-branes 
as the insertions of Wilson line and investigated 
the gluing condition by making
use of the path integral techniques.   We have further shown that 
the fuzzy sphere configuration of D0-branes directly leads to 
the Cardy states corresponding to several conjugacy classes of 
$SU(2)$ group, which finely confirms the interpretation
of spherical D2-branes as the stable bound state of D0-branes.

We also present a discussion about this subject from the view points 
of the tachyon condensation. In contrast to the flat background
any commutative configurations of D0-branes are non-BPS  
(except for the case when all the D0-branes are stacked  at one point)  
and always contain the tachyonic excitations in the open string channel.
The existence of open string tachyons implies that the 
deformation of the system is  marginally relevant, and we 
claimed that after the tachyon condensation, the system should  
flow into a fuzzy sphere configuration,
which is manifestly BPS and has no tachyons in the open string
spectrum. A similar observation was already given in \cite{alek2}. 
However, there is a subtle point in relation to our discussion.
In \cite{alek2}, being inspired by the argument of 
Kondo problem \cite{g-theorem}, the perturbation term such as  
$\dsp S_{\msc{pert}}\sim \int J^a(\sigma)S^a $ is discussed
($S^a$ should be identified with the CP matrix $M^a$ in this paper.),
and the combined currents $\hat{J}^a\equiv J^a+ S^a$ are introduced,
since they commute with the perturbation term $S_{\msc{pert}}$.
On the other hand, in our case we take the Wilson line operator 
defined by the {\em path-ordered \/} trace 
instead of $S_{\msc{pert}}$, and we have no room to consider the
combined currents like $\hat{J}^a$. In fact, the Wilson line
of fuzzy sphere actually commutes with all the currents $J^a_+$ (not
the currents like $\hat{J}^a$). It may be an important task 
to clarify the relation between these two approaches.

For the future directions it is  an interesting subject to relate 
our analysis based on the boundary conformal field theory
with the approach of low energy effective field theory, especially,
the analysis on some classical solutions of 
``unstable solitons''  \cite{hnt,hk} analogous 
to those given in the flat non-commutative spaces \cite{nc soliton}.

The analysis in the finite $k$ (or finite $N$) system 
is more challenging problem. 
However, if we intend to make the argument on the tachyon 
condensation for the D-branes in NS5 background as in section 3, 
there is a subtle point; 
the mass gap of Liouville sector in the evaluation of 
\eqn{tachyon mass} is not necessarily small in the finite $N$ case. 
We will have to carefully treat the Liouville sector to work on this 
problem.


~


\vspace{3ex}

\noindent{\Large \bf Acknowledgments}
\vspace{2ex}\\
We are grateful to M. Naka, T. Takayanagi, S. Terashima and T. Uesugi
for useful discussions. 
Y. S. also expresses his thanks to B. Pioline for valuable discussions.

The work of Y. S. is supported in part  by 
Grant-in-Aid for Encouragement of Young Scientists ($\sharp 11740142$) 
and also by Grant-in-Aid for Scientific Research on Priority Area 
($\sharp 707$) ``Supersymmetry and Unified Theory
of Elementary Particles", 
both from Japan Ministry of Education, Science, Sports and Culture.

\newpage

\section*{Appendix A ~  Some Remarks on Path Integral Representation}

\setcounter{equation}{0}
\def\theequation{A.\arabic{equation}}

In section 2 the path integral representation of Wilson line
(\ref{path integral}) was used in a formal way.
In order to remove the potential subtlety due to the UV divergence, 
we should first discretize the coordinate $\sigma$
and take the continuum limit after that.
In this appendix we define the path integral in terms of the discretized
coordinates and show that the formal analysis in section 2 can be 
confirmed with no subtlety.

We start with discretizing the coordinate $\sigma_n = n a$, 
where $a = 2\pi / N$ and $n = 1,2,\cdots,N$.
The delta function and the integral should be replaced with
\begin{equation}
 \delta (\sigma_n - \sigma_m) \to \frac{1}{a} \delta_{n,m} ~,~ 
 \oint ^{2 \pi} _{0} d\sigma \to a \sum _{n=1} ^{N} ~,
\end{equation}
and also we have
\begin{equation}
 \frac{d}{d \sigma} f (\sigma_n) \to \triangle f (\sigma_n) \equiv
  \frac{ f (\sigma_n + a) -  f (\sigma_n - a)}{2a}~.
\end{equation}     
In order to construct the action in terms of the discretized coordinates 
we have to use the dimensionless variables such as
\begin{equation}
 \hat{\triangle} = a \triangle ~,~ 
 \hat{J}^a_+ (\sigma_n) = a J^a_+  (\sigma_n) ~,~
 \hat{J}^a_- (\sigma_n) = a J^a_-  (\sigma_n) ~.
\end{equation} 

The discretized version of path integral representation should be 
defined as    
\begin{equation}
 \int \prod_{m=1}^{N} d g (\sigma_m) 
  \exp \left[ - \sum _{n=1}^{N} 
  \bra{ g (\sigma_n) } \left(\hat{\triangle} 
 + \frac{i}{k} \hat{J}^a_- (\sigma_n)
  M^a \right) \ket{ g (\sigma_n)} \right] ~,
\label{discretized path integral}
\end{equation}
where $dg(\sigma_m)$ denotes the Haar measure of $SU(2)$.
The discretized version of commutation relations of the currents 
(\ref{Jpm comm 1},\ref{Jpm comm 2}) 
\begin{eqnarray}
 {[} \hat{J}^a _{\pm} (\sigma_n) ,  \hat{J}^b _{\pm} (\sigma_m) {]} &=&
   2 \pi i \epsilon ^{ab}_{~~c} \hat{J}^c _+ (\sigma_n) \delta_{n,m}
   ~, \\
 {[} \hat{J}^a _{\pm} (\sigma_n) ,  \hat{J}^b _{\mp} (\sigma_m) {]} &=&
   2 \pi i \epsilon ^{ab}_{~~c} \hat{J}^c _- (\sigma_n) 
\delta_{n,m}
  + 2 \pi i k \delta^{ab} \frac{1}{2} (\delta_{n+1,m} - \delta_{n-1,m})~.
\end{eqnarray}
One might be afraid  that the expression \eqn{discretized path integral}
is not well-defined due to the ordering problem of  
$\hat{J}^a_{-}(\sigma_n)$ in the exponential.
However, this is not the case because of the next commutation relation
\begin{eqnarray}
 \lefteqn{{[} \hat{J}^a_- (\sigma_n) \bra{ g(\sigma_n)} 
  M^a \ket{g(\sigma_n)} ,
  \hat{J}^b_- (\sigma_m) \bra{g(\sigma_m) }M^b   \ket{g(\sigma_m) } {]}}\nn
 &=& 2 \pi i \epsilon^{abc} \hat{J}^c_+ (\sigma_n) 
 \bra{g(\sigma_n)} M^a \ket{ g(\sigma_n)} 
\bra{ g(\sigma_n)} M^b \ket{ g(\sigma_n)}  
 \delta_{n,m} \nn
 &=& 0 ~.
\label{JM comm}
\end{eqnarray}
Since the coordinates are discrete,
we can use Kronecker symbol rather than Dirac $\delta$-function,
which removes the subtlety of the divergence of equal $\sigma$ 
$\delta$-function. In this way we  conclude that the pass integral
\eqn{discretized path integral} is well-defined with no problem
of the operator ordering.

We also remark that we need not here take account of the gauge fixing of
$U(1)$-gauge symmetry: $g(\sigma_m)~ \longrightarrow~
g(\sigma_m)a(\sigma_m) $, $(\forall a(\sigma_m) \in U(1))$ 
for the discretized framework. The gauge volume is finite as in 
the usual lattice gauge theory.



In our argument of section 2 it is quite important that 
the Wilson line of fuzzy sphere configuration \eqn{fuzzy CP} 
commutes with the currents $J^a_+ (\sigma)$. 
We now show that this is indeed the case for the discretized  
version of Wilson line \eqn{discretized path integral}.
For this purpose we only have to 
replace the unitary operators (\ref{UU}) with
\begin{equation}
 \hat{U} = \exp 
 \left( i \sum_{n=1}^N u^a (\sigma_n) \hat{J}_+ ^a (\sigma_n) \right) ~,~~
 U(\sigma_n) =  \exp ( - 2 \pi i  u^a (\sigma_n) M^a)~.
\end{equation}
Then the following identities  
\begin{equation}
  \hat{U} \left[ \hat{\triangle} 
  + \frac{i}{k} \hat{J}^a_- (\sigma_n) M^a \right]\hat{U}^{-1} = 
 U(\sigma)  \left[ \hat{\triangle} + 
  \frac{i}{k} \hat{J}^a_- (\sigma_n) M^a \right]U(\sigma)^{-1}~,
\end{equation}
are also satisfied as in \eqn{DD}.
Thus we can likewise show that the Wilson line commutes with the currents 
$J^a_+(\sigma)$ by using the identity;
  \begin{eqnarray}
 \lefteqn{\hat{U} \int \prod_{m=1}^{N} d g (\sigma_m) 
  \exp \left[ - \sum _{n=1}^{N} 
  \bra{ g (\sigma_n) }\left( \hat{\triangle} + 
   \frac{i}{k} \hat{J}^a_- (\sigma_n)
  M^a \right) \ket{ g (\sigma_n)} \right] \hat{U}^{-1}} \nn
    &=& 
   \int \prod_{m=1}^{N} d g (\sigma_m) 
  \exp \left[ - \sum _{n=1}^{N} 
  \bra{ g (\sigma_n) } U(\sigma_n) \left(
  \hat{\triangle} + \frac{i}{k} \hat{J}^a_- (\sigma_n)
  M^a  \right) U(\sigma_n)^{-1}\ket{ g (\sigma_n)} \right]
\end{eqnarray} 
and the fact that the Haar measure $dg(\sigma_n)$ is 
invariant under the field redefinition 
$g(\sigma_n) \to  g'(\sigma) =  U (\sigma_n)^{-1}g(\sigma_n)$.

Up to now, we investigated  the properties of the path integral
representation when the coordinate $\sigma$ is discretized.
We have to take the continuum limit $a \, \rightarrow\, 0$
to define the Wilson line operator (\ref{path integral}) in section 2.  
Since it generically has divergent contributions 
in this limit, we will have to take account of the renormalization of 
coupling constants  and would potentially suffer non-trivial 
radiative corrections. However, in our case 
\eqn{discretized path integral}, the story becomes quite simple
as long as it is inserted at the boundary with the suitable
gluing condition \eqn{glue1}, of which discretized version is
\begin{equation}
\hat{J}^a_+(\sigma_n)\ket{B}=0~.
\label{dglue1}
\end{equation}
We showed in the above argument that the Wilson line operator 
\eqn{discretized path integral} preserves 
the symmetry of discretized currents $\hat{J}_+^a(\sigma_n)$, 
and thus we obtain
\begin{equation}
\hat{J}^a_+(\sigma_n) \, W(\{M^a\};a) \ket{B}=0~,
\label{dJWB}
\end{equation}
where $ W(\{M^a\};a)$ denotes the Wilson line operator defined in 
\eqn{discretized path integral}.
 
The discretized conformal invariance at boundary should 
be realized in terms of the discretized boundary stress tensor;
\begin{equation}
\hat{T}_-(\sigma_n) 
= \frac{1}{k}\hat{J}_-^a(\sigma_n)\hat{J}_+^a(\sigma_n)
\equiv \sum_{m\in \bsz_N}\, \hat{L}_{-,\,m}e^{-im\sigma_n}~,
\label{dbemt}
\end{equation}  
which is the discretized counterpart of \eqn{bemt} as we will discuss below.
Notice that the products of currents $\hat{J}^a_{\pm}(\sigma_n)$
at the equal points
are now well-defined without any subtlety of UV divergence.  
We can directly check that the mode oscillators 
$\hat{L}_{-,\,m}$ generate a closed algebra together with 
$\hat{J}^a_{\pm,\,n}$ 
(defined by $\dsp \hat{J}^a_{\pm}(\sigma_l)
=\sum_{n\in \bsz_N} \hat{J}^a_{\pm,\,n} e^{-i n\sigma_l}$);
\begin{eqnarray}
\lb \hat{L}_{-,\,m},\, \hat{J}^a_{\pm,\,n} \rb &=& 
-\frac{1}{a}\sin\left(an\right)\, \hat{J}^a_{\pm,\,n+m}~,
\nonumber \\
\lb \hat{L}_{-,\,m},\, \hat{L}_{-,\,n} \rb &=& 
\frac{1}{a} \left\{\sin\left(am\right)-
\sin\left(an \right)\right\} \hat{L}_{-,\, m+n} \nonumber\\
&&\hspace{2cm}
+ \sum_{l\in \bsz_N}\, c(m,n,l;a)\hat{J}^a_{-,\,m+n-l}\hat{J}^a_{+,\,l}~,
\label{comm d Virasoro} 
\end{eqnarray} 
where $c(m,n,l;a)$ are some constants depending on $m,n,l\in \bz_N$ 
and are of order $a$.

One might think our definition \eqn{dbemt} to be peculiar, since
we include the factor $1/k$ rather than the usual one $1/(k+2)$.
By our construction of discretized currents we need not introduce
the normal ordering, and the absence of the level 
shift $k\,\rightarrow\, k+2$ is originating from this fact.
By this reason it seems subtle whether the continuum limit
of $\hat{L}_{-,\,n}$ truly corresponds to the mode oscillator of 
boundary stress tensor $L_n-\tilde{L}_{-n}$ \eqn{bemt} 
(defined with the usual normal ordering).
However, the commutation relations 
\eqn{comm d Virasoro} imply  that the continuum limits\footnote{
In taking $a$ to zero limit, a subtle point is in the zero-mode part, 
since the normal ordering contribution would become important.
However,  we can show that
$\bra{0}\hat{L}_{-,\,0}\ket{0}= 0$ and $
\bra{0}\lb \hat{L}_{-,\,n},\,\hat{L}_{-,\,-n} \rb\ket{0}=0$
hold for arbitrary $a$ without taking the normal ordering.
The essential point is the cancellation of contributions from 
the central terms  of left and right moving sectors
and the second equality is derived from the property
$c(n,-n,l;a)=c(n,-n,-l;a)$.
} of $\hat{L}_{-,\,n}$ satisfy the same
commutation relations with the currents $J^a_{\pm,\,n}$ as those of
$L_n-\tilde{L}_{-n}$, and hence they are identified with each other  
{\em on the states of the type;}
$$
\dsp \sum_{\stackrel{a=\{a_1,a_2,\ldots\}}{n=\{n_1,n_2,\ldots\}},\,r}
\cN_{a,n,r}\prod_{i}J^{a_i}_{-,\,n_i}
\ket{B;r}~,
$$
where $\ket{B;r}$ satisfies \eqn{dglue1}.
All the states considered in relation to the Wilson lines in this paper
are indeed of this type.
Therefore we can regard \eqn{dbemt} as the discretized version 
of \eqn{bemt} in our arguments.

Equation \eqn{dJWB} and the definition \eqn{dbemt} readily implies 
\begin{equation}
\hat{T}_-(\sigma_n)\, W(\{M^a\};a)\ket{B}=0~.
\end{equation}
In this way we can conclude that the boundary state
$ W(\{M^a\};a)\ket{B}$ preserves the discretized conformal invariance
{\em for an arbitrary  finite lattice spacing $a$}.
This fact means that $W(\{M^a\};a)\ket{B}$ corresponds 
to the fixed point of boundary renormalization group flow\footnote
     {This aspect seems to be consistent with the perturbative 
     calculation of $\beta$-function presented in the works 
     \cite{g-theorem}.}, 
and thus we can take the continuum limit without suffering 
the renormalization and any radiative corrections.
Therefore we can safely conclude that the Wilson line operator
in the continuous theory (\ref{path integral})
(with the CP matrices \eqn{fuzzy CP}) 
preserves the gluing condition \eqn{glue1}.
That is truly the statement we need for our argument in section 2.

For a general Wilson line operator such as 
$ \tr \left(P\exp \left(i\la  \dsp \oint J^a_- (\sigma)M^a \right)\right)$
(namely, with a  general $\la$, and general matrices $M^a$), 
taking the continuum limit will be of course  a  non-trivial problem 
with complicated radiative corrections, since it is not a 
truly marginal operator. This fact makes the rigid analysis 
away from the conformal points difficult and it is beyond 
the scope of this paper.


\section*{Appendix B ~ Extended Supersymmetry}
\setcounter{equation}{0}
\def\theequation{B.\arabic{equation}}

In this appendix  we show that our boundary state defined in 
the CHS background  actually
preserves the $N=2$ and $N=4$ superconformal symmetries. 
In addition to the $SU(2)$ supercurrents defined in section 3,
the Liouville mode is expressed as $\phi$ 
and its superpartner as $\psi^{\phi}$, which has
the OPE;\footnote{We here use the different normalization of fermions in
$SU(2)$ sector and their OPEs are given by
$$ \psi^a (z) \psi^b (0) \sim \frac{\delta^{ab}}{z}.$$}  
$\psi^{\phi}(z)\psi^{\phi}(0)\sim 1/z$.
We also define
\begin{equation}
 \psi^{\pm} = \frac{1}{\sqrt{2}}(\psi^1 \pm i \psi^2)~,~~
 \Psi^{\pm} = - \frac{1}{\sqrt{2}}(\psi^3 \pm i \psi^{\phi})~,~~
 j^{\pm} = j^1 \pm i j^2~.
\end{equation}

First, we investigate the system as  an  $N=2$ superconformal field theory.
The realizations of the $N=2$ superconformal currents are given by
\begin{eqnarray}
 T &=& - \frac{1}{2} (\partial \phi)^2 - \frac{Q}{2} \partial^2 \phi
  + \frac{1}{N}(j^a j^a) - \frac{1}{2}(\psi^a \partial \psi^a) 
- \frac{1}{2}(\psi^{\phi} \partial \psi^{\phi})   ~, \nn
 G^{\pm} &=& - \frac{1}{\sqrt{2}}
  \left(\sqrt{\frac{2}{N}} J^3 \pm \partial \phi\right)
   \Psi^{\pm} + \frac{1}{\sqrt{N}}(j^{\pm} \psi^{\mp}) 
   \mp \frac{Q}{\sqrt{2}} \partial \Psi ^{\pm} ~, \nn
 J &=& \Psi^+ \Psi^- - \psi^+ \psi^-~,
\label{N=2 SCA}
\end{eqnarray}
where $a=1,2,3$ and we will use these indices below.
In this theory two types of the boundary conditions preserving 
the $N=2$ SUSY are possible  
and called the A-type and the  B-type conditions 
\cite{ooy};

\noindent{\underline{\bf A-type}}
 \begin{eqnarray}
 (J(\sigma) - \tJ(\sigma)) \ket{B;\ep} &=& 0 ~, \nn
 (G^{\pm}(\sigma) - i \ep \tG^{\mp}(\sigma)) \ket{B;\ep} &=& 0~, 
\label{a-type}
\end{eqnarray}
\noindent{\underline{\bf B-type}}
  \begin{eqnarray}
 (J(\sigma) + \tJ(\sigma)) \ket{B;\ep} &=& 0 ~, \nn
 (G^{\pm}(\sigma) - i \ep \tG^{\pm}(\sigma)) \ket{B;\ep} &=& 0~. 
\label{b-type}
\end{eqnarray}

The gluing conditions compatible with the B-type boundary condition
\eqn{b-type} are  given by
\begin{eqnarray}
 (j^{\pm}(\sigma) + e^{\pm i \alpha} \tj^{\pm}(\sigma)) \ket{B;\ep} &=& 0 ~,
\nn
 (j^3(\sigma) + \tj^3(\sigma)) \ket{B;\ep} &=& 0 ~, \nn
 (\partial \phi(\sigma) + \bar{\partial}\tphi(\sigma)-Q) \ket{B;\ep} 
  &=& 0 ~, \nn
 (\psi^{\pm}(\sigma) + i \ep e^{\pm i \alpha} \tpsi^{\pm}(\sigma)) 
    \ket{B;\ep} &=& 0~, \nn
 (\psi^{3,\phi}(\sigma) + i \ep  \tpsi^{3,\phi}(\sigma)) \ket{B;\ep}
    &=& 0 ~,
\label{n=1free}
\end{eqnarray}  
where $\alpha \in \br$ and there is the  momentum 
shift $Q$ in  the Liouville mode
(see, for example,  \cite{eguchi}). 
The case of $\alpha =0$ is used in section 3 to define the Cardy states
\eqn{s Cardy}.
It is obvious that the Wilson line of fuzzy sphere does not break
the $N=2$ SUSY. For the commutative CP matrices, $M^1=M^2=0$ and 
$M^3=\mbox{diag}(a_1,a_2,\ldots)$, let us consider 
the decomposition like \eqn{s bs comm}.
For the each term we obtain the gluing conditions \eqn{n=1free}
with the various phase $\al$ depending on the value $a_i$.
Hence the boundary state with such Wilson line obeys the B-type 
condition as well.
In this way we have shown that the $N=2$ 
superconformal symmetry is surely preserved on
the boundary states we considered in section 3.

On the other hand, we have to change the gluing conditions to make it 
compatible with the A-type condition as follows;
  \begin{eqnarray}
 (j^{\pm}(\sigma) + e^{\pm i \alpha} \tj^{\mp}(\sigma)) \ket{B;\ep} &=& 0 
  ~, \nn
 (j^3(\sigma) - \tj^3(\sigma)) \ket{B;\ep} &=& 0~, \nn
 (\partial \phi(\sigma) + \bar{\partial}\tphi(\sigma)-Q) \ket{B;\ep} 
&=& 0 ~, \nn
 (\psi^{\pm}(\sigma) + i \ep e^{\pm i \alpha} \tpsi^{\mp}(\sigma))
    \ket{B;\ep} &=& 0 ~, \nn
 (\psi^{3}(\sigma) - i \ep  \tpsi^{3}(\sigma)) 
      \ket{B;\ep} &=& 0 ~, \nn
 (\psi^{\phi}(\sigma) + i \ep  \tpsi^{\phi}(\sigma)) 
      \ket{B;\ep} &=& 0 ~,
\end{eqnarray}  
which is an example of general gluing condition \eqn{glue0}.

Next, we focus on the (small) $N=4$ superconformal structure.
This theory has  $SU(2)_R$ currents $A^a$
and two more superconformal currents other than $N=2$ ones. 
We take their linear combinations so that
one is a singlet under the $SU(2)$ transformation and the others are
vectors. 
Their explicit forms in the CHS $\sigma$-model
are  obtained  as follows \cite{sevrin,chs};
\begin{eqnarray}
 T &=& - \frac{1}{2} (\partial \phi)^2 - \frac{Q}{2} \partial^2 \phi
  + \frac{1}{N}(j^a j^a) - \frac{1}{2}(\psi^a \partial \psi^a) 
- \frac{1}{2}(\psi^{\phi} \partial \psi^{\phi}) ~, \nn
 G^0 &=& i \partial \phi \psi^{\phi} + Q i \partial \psi^{\phi} 
  + \sqrt{\frac{2}{N}}(j^a \psi^a - i \psi^1 \psi^2 \psi^3 )~,  \nn
  G^a &=& i \partial \phi \psi^a + Q i \partial \psi^a 
  + \sqrt{\frac{2}{N}}(- j^a \psi^{\phi} + \epsilon^{abc}j^b \psi^c 
  + \frac{i}{2}\epsilon^{abc} \psi^{\phi} \psi^b \psi^c )  ~, \nn
 A^a &=& -\frac{i}{2}\psi^{\phi} \psi^a 
  - \frac{i \epsilon^{abc}}{4}\psi^b \psi^c ~.
\label{N=4 SCA}
\end{eqnarray} 

In general  the  boundary  condition preserving the $N=4$ SUSY is
given by \cite{ooy};
\begin{eqnarray}
  (A^a(\sigma) + \Lambda^{a}_{~b} \tA^b(\sigma)) \ket{B;\ep} &=& 0~, \nn
 (G^0(\sigma) - i \ep \tG^0(\sigma)) \ket{B;\ep} &=& 0 ~, \nn 
  (G^a(\sigma) - i \ep \Lambda^{a}_{~b}\tG^b(\sigma)) \ket{B;\ep} &=& 0~, 
\label{N=4 bc}
\end{eqnarray}
where $\Lambda^{a}_{~b}$ is an automorphism of $SU(2)$.
This condition is compatible with the twisted  gluing condition of 
the type \eqn{glue0};
  \begin{eqnarray}
 (j^a(\sigma) +\Lambda^{a}_{~b} \tj^b(\sigma)) \ket{B;\ep} &=& 0~, \nn
 (\partial \phi(\sigma) + \bar{\partial}\tphi(\sigma)-Q) \ket{B;\ep} 
 &=& 0 ~, \nn
 (\psi^a(\sigma) + i \ep \Lambda^{a}_{~b} \tpsi^b(\sigma)) 
      \ket{B;\ep} &=& 0 ~, \nn
 (\psi^{\phi}(\sigma) + i \ep  \tpsi^{\phi}(\sigma)) 
      \ket{B;\ep} &=& 0~,
\end{eqnarray}  
and thus we can easily construct the boundary state satisfying 
the $N=4$ boundary condition \eqn{N=4 bc}. 
Notice that the gluing condition (\ref{n=1free}) with $\alpha =0$ 
is the  special case with
$\Lambda^{a}_{~b} = \delta^{a}_{~b}$ and hence the boundary states 
\eqn{s Cardy} satisfy this condition. More general cases  with 
the Wilson lines can be also discussed just as in the $N=2$ argument.

~


\section*{Appendix C ~ Convention of Conformal Field Theory}

\setcounter{equation}{0}
\def\theequation{C.\arabic{equation}}

\subsection*{1. Theta functions}

The Jacobi theta functions are defined by
\begin{eqnarray}
&&\theta_1(q,z)=i\sum_{n=-\infty}^{\infty}
(-1)^nq^{\frac{1}{2}\left(n-\frac{1}{2}\right)^2}z^{n-\frac{1}{2}}~,\quad
\theta_2(q,z)=\sum_{n=-\infty}^{\infty}
q^{\frac{1}{2}\left(n-\frac{1}{2}\right)^2}z^{n-\frac{1}{2}}~, \nonumber\\
&&\theta_3(q,z)=\sum_{n=-\infty}^{\infty}q^{\frac{n^2}{2}}z^n~,
\quad\quad\quad\quad\quad\quad\quad
\theta_4(q,z)=\sum_{n=-\infty}^{\infty}(-1)^n
q^{\frac{n^2}{2}}z^n ~,
\label{jacobi}
\end{eqnarray}
where we set $q=e^{2\pi i \tau}$ and $z=e^{2\pi i \nu}$. For an arbitrary
positive integer $k$, the  theta function of level $k$ is defined by
\begin{eqnarray}
\theta_{m,k}(q,z)&=&\sum_{n=-\infty}^{\infty}
q^{k\left(n+\frac{m}{2k}\right)^2}
z^{k\left(n+\frac{m}{2k}\right)}~,\\
m&=&-k+1,\ldots ,k~.\n
\end{eqnarray}
Therefore we can rewrite the Jacobi theta functions in terms of the theta
function of level 2:  
\begin{eqnarray}
i\theta_1(q,z)=\theta_{1,2}(q,z)-\theta_{3,2}(q,z)~,\qquad &&
\theta_2(q,z)=\theta_{1,2}(q,z)+\theta_{3,2}(q,z) ~,\n\\
\theta_3(q,z)=\theta_{0,2}(q,z)+\theta_{2,2}(q,z)~,\qquad &&
\theta_4(q,z)=\theta_{0,2}(q,z)-\theta_{2,2}(q,z)~.
\end{eqnarray}

\subsection*{2. Characters of $N=2$ minimal model}

There is a discrete series of unitary representations of $N = 2$
superconformal algebra with
$c <3$, namely, with  $\dsp c={3k \over k+2}$ ($k=1,2,3,\ldots$).
Based on these representations one can construct the family  
of  rational conformal field theories known as the $N = 2$ minimal models.
The discrete representations of the  $N=2$ algebra are 
related to the ${\widehat {SU(2)}}_k$ representations.
The character of ${\widehat {SU(2)}}_k$ with the spin ${\ell\over 2}$ 
($0\leq \ell\leq k$) representation is calculated as 
\begin{equation}
\chi_{\ell}^{(k)}(q)=\frac{\theta_{\ell+1,k+2}-\theta_{-\ell-1,k+2}}
{\theta_{1,2}-\theta_{-1,2}}(q)
:=\sum_{m\in{\bf Z}_{2k}}c_m^{\ell}(q)\theta_{m,k}(q)~,
\end{equation}
and the coefficient $c_m^{\ell}(q)$ is called the string function. 
The character of $N=2$ representation labeled by $(\ell, m, s)$ 
is obtained through the ``branching relation'' \cite{gepner};
\begin{equation}
\chi_{\ell}^{(k)}(q)\theta_{s,2}(q)=\sum_{m=-k-1}^{k+2}
\chi^{\ell,s}_m(q)\theta_{m,k+2}(q)~,
\label{branching relation}
\end{equation}
where we set
\begin{equation}
\chi_{m}^{\ell,s}(q)=
\sum_{r \in {\bf Z}_k} c_{m-s+4r}^{\ell}(q)\theta_{2m+(k+2)(-s+4r),2k(k+2)}
\left(q\right) ~.
\end{equation}
These ``branching functions'' $\chi_{m}^{\ell,s}(q)$
are defined in the range $\ell\in\{0,\ldots,k\}$, 
$m\in Z_{2k+4}$, $s\in Z_4$ and $\ell +m+s=0$ mod 2. 
The characters of $N=2$ minimal model of level $k$ are 
then expressed  as 
\begin{eqnarray}
&&\C_{\ell m}^{(NS)}(q)
=\chi_m^{\ell,0}(q)+\chi_m^{\ell,2}(q)~,\quad
\widetilde{\C}_{\ell m}^{(NS)}(q)
=\chi_m^{\ell,0}(q)-\chi_m^{\ell,2}(q)~,\n\\
&&\C_{\ell m}^{(R)}(q)
=\chi_m^{\ell,1}(q)+\chi_m^{\ell,3}(q)~,\quad
\widetilde{\C}_{\ell m}^{(R)}(q )
=\chi_m^{\ell,1}(q)-\chi_m^{\ell,3}(q)~.
\end{eqnarray}


\end{document}